\documentclass{aa}
\usepackage{graphicx,aalongtable}
\usepackage{txfonts}
\begin{document}
\title{Abundance analysis of barium and mild barium stars
\thanks{Observations collected at ESO, La Silla, Chile, within the ON/ESO agreements.}}
%\subtitle{}
\author{
R. Smiljanic\inst{1}\fnmsep\thanks{Present address:
Universidade de S\~ao Paulo, IAG, Dpt. de Astronomia, Rua do Mat\~ao 1226, S\~ao Paulo-SP 05508-900, Brazil}
\and
G.F. Porto de Mello\inst{1}
\and
L. da Silva\inst{2}
 }
\offprints{R. Smiljanic}

\institute{Observat\'orio do Valongo, Universidade Federal do Rio de Janeiro, Ladeira do Pedro Ant\^onio 43, Sa\'ude, Rio de Janeiro-RJ 20080-090, Brazil\\
\email{rodolfo@astro.iag.usp.br;gustavo@ov.ufrj.br}
\and
Observat\'orio Nacional, Rua Gal. Jos\'e Cristino 77, S\~ao Cristov\~ao, Rio de Janeiro-RJ 20921-400, Brazil\\
\email{licio@on.br}
 }
\date{Received / Accepted}

% \abstract{}{}{}{}{}
% 5 {} token are mandatory

  \abstract
  % context heading (optional), leave it empty if necessary
{}
   {We compare and discuss
   abundances and trends in normal giants, mild barium, and barium
   stars, searching for differences and similarities between barium
   and mild barium stars that could help shed some light on the origin
   of these similar objects. Also, we search for nucleosynthetic
   effects possibly related to the s-process that were observed in the
   literature for elements like Cu in other types of s-process enriched
   stars.}
   {High signal to noise, high resolution spectra were obtained for a sample
   of normal, mild barium, and barium giants. Atmospheric parameters were
   determined from the \ion{Fe}{i} and \ion{Fe}{ii} lines. Abundances for Na,
   Mg, Al, Si, Ca, Sc, Ti, V, Cr, Mn, Fe, Co, Ni, Cu, Zn, Sr, Y, Zr, Ba, La,
   Ce, Nd, Sm, Eu, and Gd, were determined from equivalent widths and model
   atmospheres in a differential analysis, with the red giant $\epsilon$ Vir
   as the standard star.}
   {The different levels of s-process overabundances of barium and mild barium
   stars were earlier suggested to be related to the stellar metallicity.
   Contrary to this suggestion, we found in this work no evidence of barium
   and mild barium having a different range in metallicity. However, comparing
   the ratio of abundances of heavy to light s-process elements, we found some
   evidence that they do not share the same neutron exposure parameter. The
   exact mechanism controlling this difference is still not clear. As a by-product
   of this analysis we identify two normal red giants misclassified as mild barium
   stars. The relevance of this finding is discussed. Concerning the suggested
   nucleosynthetic effects possibly related to the s-process, for elements like
   Cu, Mn, V and Sc, we found no evidence for an anomalous behavior in any of
   the s-process enriched stars analyzed here. However, further
   work is still needed since a clear [Cu/Fe] vs. [Ba/H] anticorrelation exists
   for other s-process enriched objects.}
   {}


\maketitle
%
%________________________________________________________________

%________________________________ INTRODUCTION ______________________
%--- 1 ---

\section{Introduction}

Barium stars are chemically peculiar G$-$K giants first identified
by Bidelman \& Keenan (\cite{BK51}). These stars were found to
have the Ba II 4554 \AA\ resonance line, the CH G band and the Sr
II 4077 \AA\ and 4215 \AA\ lines abnormally enhanced. Abundance
analyses (Burbidge \& Burbidge \cite{BB57}; Warner \cite{W65}) 
confirmed that such features were due to real
atmospheric overabundances of carbon and of the heavy s-process
elements.

S-process nucleosynthesis itself is only expected in thermally
pulsing asymptotic giant branch (AGB) stars. Two reactions are the
main providers of neutrons, $^{22}$Ne($\alpha$,n)$^{25}$Mg and
$^{13}$C($\alpha$,n)$^{16}$O. Although the $^{22}$Ne reaction was
earlier thought to be dominant, it was shown (Tomkin \& Lambert
\cite{TL79}; Tomkin \& Lambert \cite{TL83}; Malaney \cite{M87a})
that it produces results incompatible with the observations. Thus,
the most important neutron source is now thought to be the
$^{13}$C reaction. The source of $^{13}$C, however, is not well
established. Recent works on s-process enrichment usually
parameterize the amount of $^{13}$C burnt during the s-process
operation (Busso et al.\@ \cite{B95}, \cite{B01}; Gallino et al.\@
\cite{G98}).

In AGBs, the deep dredge-up phenomena that follows the thermal pulses, the
so-called third dredge-up, mixes some of the
processed material to the atmosphere, where it becomes accessible to observations.
Barium stars, however, are less massive and less luminous than the
AGB stars. It is not expected for barium stars to synthetize s-process elements in
their interiors or even to be able to dredge this material up to the surface,
thus they cannot be self-enriched.

An important discovery that led to the solution of this problem
was made in the early eighties. Through the radial velocity
monitoring of a sample of barium and normal giants, it was
discovered that all barium giants are likely members of binary
systems (McClure et al.\@ \cite{MFN80}; McClure \cite{Mc83},
\cite{Mc84}). White dwarf companions were also detected in the UV
with the International Ultraviolet Explorer (IUE) (B\"ohm-Vitense
\cite{BV80}; Domini \& Lambert \cite{DL83}; B\"ohm-Vitense \&
Johnson \cite{BJ85}). More recent radial velocities monitoring
(Udry et al.\@ \cite{U98a}, \cite{U98b}) and UV observations
(B\"ohm-Vitense et al.\@ \cite{BV00}) with the Hubble Space
Telescope (HST) have supported the idea of binarity.

\begin{table*}
\caption{Data of the sample stars. Visual magnitudes and spectral
types are from the SIMBAD database. The effective temperatures
(T$_{\rm eff}$), log g, metallicities and barium abundances are
from Zacs (\cite{Z94}) for all stars with the exception of \object{HR 1016}
(Pilachowski \cite{P77}), \object{HR 4932} (McWilliam \cite{M90}), and \object{HR
5058} (Luck \& Bond \cite{LB91}). The colors (V$-$K) and (R$-$I)
are from Hoffleit \& Jaschek (\cite{BSC}) and Johnson
(\cite{J66}). The stars \object{HR 440}, \object{HR 1326} and HR 4932 ($\epsilon$
Vir) are the three normal giants included in the sample.}
\label{tab:1} \centering
\begin{tabular}{c c c c c c c c c c c}
\noalign{\smallskip}
\hline\hline
\noalign{\smallskip}
HR & HD & $V$ & (V$-$K) & (R$-$I) & ST & T$_{\rm eff}$ & log g & [Fe/H] & [Ba/Fe] \\
\hline
440 & 9362 & 4.0 & -- & +0.51 & K0III-IV    & -- & -- & -- & -- \\
649 & 13611 & 4.4 & +2.06 & +0.49 & G6II-III & 5050 & 2.3 & $-$0.3 & +0.44 \\
1016 & 20894 & 5.5 & -- & +0.47 & G6.5IIb & 5100 & 3.6 & $-$0.2 & $-$0.20 \\
1326 & 26967 & 3.9 & +2.48 & +0.59 & K1III & -- & --  & -- & -- \\
2392 & 46407 & 6.2 & +2.21 & +0.48 & K0III & 5000 & 2.1 & +0.1 & +1.34 \\
4608 & 104979 & 4.1 & +2.22 & +0.49 & G8III & 5000 & 2.2 & $-$0.1 & +0.93 \\
4932 & 113226 & 2.8 & +2.04 & +0.45 & G8IIIab & 5060 & 2.97 & +0.15 & -- \\
5058 & 116713 & 5.1 & -- & -- & K0.5III & 5000 & 3.0 & +0.2 & -- \\
5802 & 139195 & 5.3 & -- & +0.45 & K0III & 5140 & 2.7 & +0.3 & +0.52 \\
7321 & 181053 & 6.4 & -- & -- & K0IIIa & 4885 & 2.1 & $-$0.2 & +0.34 \\
8115 & 202109 & 3.2 & +2.11 & +0.48 & G8.5III & 5050 & 2.8 & +0.1 & +0.41 \\
8204 & 204075 & 3.7 & +1.88 & +0.43 & G4II & 5230 & 1.5 & +0.2 & +1.31 \\
 -- & 205011 & 6.4 & -- & -- & G9IIIa & 4950 & 2.4 & +0.1 & +0.88 \\
8878 & 220009 & 5.1 & -- & -- & K2III & 4575 & 2.6 & $-$0.1 & +0.42 \\
\noalign{\smallskip}
\hline
\end{tabular}
\end{table*}

The chemical peculiarities of the barium stars are then directly
related to their binarity through a mass transfer scenario. The
companion star seen today as a white dwarf was initially the more
massive star of the system. As such it evolved faster and became a
thermally pulsing AGB, while the current barium star was still in
an earlier evolutionary stage. The star then enriched its He
burning envelope with products of the s-process nucleosynthesis
and through successive third dredge-ups mixed this material to the
atmosphere. This enriched material was then transferred onto the
current barium star through mass loss mechanisms (Jorissen \&
Mayor \cite{JM92}; Liang et al.\@ \cite{L00}). Thus, the
overabundances are not intrinsic to the barium star but are
important observational tests of the theories of nucleosynthesis,
convection and mass loss in cool stars and the study of the
chemical evolution of the Galaxy.

Concerning the details of the s-process nucleosynthesis, there is
some observational evidence pointing towards a possible
preferential depletion of certain iron peak elements. Main
sequence stars of the young Ursa Major moving group (UMaG
hereafter), which are s-process enriched, show a Cu depletion with
respect to Fe (Castro et al.\@ \cite{C99}). Other barium enriched
objects show the same behavior (Pereira \& Porto de Mello
\cite{P97}, Pereira et al.\@ \cite{P98}). Particularly, the
abundance pattern of \object{HR 6094} (Porto de Mello \& da Silva
\cite{PS97}), an UMaG member, suggests the depletion of Mn and Cu
and that V and Sc could have been preserved with respect to Fe.
Such results may represent important constraints to the neutron
capture models in AGBs and deserve further investigation. This
kind of data is still very scarce in the literature.

Detailed abundance analyses of barium
(or mild barium) stars with modern high quality data are scarce in
the literature (Boyarchuk et al.\@ \cite{B02}; Liang et al.\@
\cite{Li03}; Antipova et al.\@ \cite{A04}; Yushchenko et al.\@
\cite{Y04}; Allen \& Barbuy \cite{AB06}). Most of the available analyses are based on data with
lower S/N ratio (Pilachowski \cite{P77}, Smith \cite{S84}, Kovacs
\cite{K85}, Luck \& Bond \cite{LB91}, Zacs \cite{Z94}). Thus,
abundance errors in these works could possibly be blurring the
small scale of the suggested nucleosynthetic effects.

In this work we derived atmospheric parameters and the detailed
chemical composition of a sample containing eleven stars
classified in the literature as barium or mild barium stars and
three normal giants for comparison purposes. Abundances were
obtained for the light elements Na, Mg, Al, Si, Ca; the iron peak
elements Sc, Ti, V, Cr, Mn, Fe, Co, Ni; Cu, Zn (considered as
transition elements between the iron peak and s-process elements);
the s-process elements Sr, Y, Zr, Ba, La, Ce, Nd; and the
r-process dominated elements Sm, Eu, Gd.

The abundances of the barium and mild barium stars are compared
and the possible nucleosynthetic effects discussed. In addition,
we discuss the relevance of the identification in this work of two
normal giants previously misclassified as mild barium stars. Such
a result shows the necessity of high quality analysis for this
class of peculiar stars. The observations are described in Sect.
2, the stellar parameters in Sect. 3 and the abundances in Sect.
4. In Sect. 5 we discuss the results and the nucleosynthetic
effects while conclusions are drawn in Sect. 6.

%____________________________ OBSERVATIONS ___________________________
%--- 2 ---
\section {Observational data}

High resolution CCD spectra were obtained using the FEROS (Kaufer et al.\@
\cite{Kau99}) spectrograph at the ESO 1.52 m telescope at La Silla, Chile. FEROS
is a fiber-fed echelle spectrograph that provides a full
wavelength coverage of $\lambda$$\lambda$3500$-$9200 over 39
orders at a resolving power of R = 48000. The detector used was an
EEV CCD chip with 2048 x 4096 pixels and a pixel size of 15
$\mu$m. All spectra were reduced using the FEROS pipeline
software. The spectra have a typical signal to noise ratio of S/N
$\approx$ 500$-$600.

The sample stars were selected based on previous literature
analyses. It was a concern to include stars with different
overabundances, ranging from mild barium stars to barium stars
with overabundances larger than 1.0 dex. It is important to span
such a variety of barium stars in order to search for possible
correlations among the observed overabundances and the atmospheric
and evolutionary parameters. Table \ref{tab:1} presents the
literature data on the selected objects. All the stars, with the
exception of HR 1016 (Pilachowski \cite{P77}) and HR 5058 (Luck \&
Bond \cite{LB91}), were analyzed by Zacs (\cite{Z94}) who
conducted a detailed analysis of a larger number of barium stars
but used data with much lower S/N than ours.
\begin{table*}
\caption{The atmospheric parameters, effective temperature
(T$_{\rm eff}$), surface gravity (log g), microturbulence velocity
($\xi$) and metallicity ([Fe/H]) (we use throughout the notation
[A/B] = log (N(A)/N(B)$_{\rm star}$ - log (N(A)/N(B)$_{\rm
\odot}$), derived for the sample stars as described in the text.
The values for [\ion {Fe}{i}/H] and [\ion {Fe}{ii}/H] are followed
by the standard deviation and the number of lines on which the
abundance is based.} \label{tab:par} \centering
\begin{tabular}{c c c c c c c c}
\noalign{\smallskip}
\hline\hline
\noalign{\smallskip}
Star & T$_{\rm eff}$ (K) & log g & $\xi$(km/s) & [\ion{Fe}{i}/H] $\pm$ $\sigma$\,(\#)& [\ion{Fe}{ii}/H] $\pm$ $\sigma$\,(\#) \\
\hline
HR 440   & 4780 & 2.43 & 1.71 & $-$0.34 $\pm$ 0.07 (129) & $-$0.33 $\pm$ 0.07 (13) \\
\object{HR 649}   & 5120 & 2.49 & 1.96 & $-$0.14 $\pm$ 0.06 (117) & $-$0.15 $\pm$ 0.07 (13) \\
HR 1016 & 5080 & 2.60 & 1.80 & $-$0.11 $\pm$ 0.06 (119) & $-$0.11 $\pm$ 0.09 (13) \\
HR 1326 & 4650 & 2.51 & 1.52 & $+$0.00 $\pm$ 0.07   (119) & $+$0.01 $\pm$ 0.16 (13) \\
\object{HR 2392} & 4940 & 2.65 & 1.87 & $-$0.09 $\pm$ 0.12 (115) & $-$0.08 $\pm$ 0.15 (12) \\
\object{HR 4608} & 4920 & 2.58 & 1.71 & $-$0.35 $\pm$ 0.05 (117) & $-$0.34 $\pm$ 0.06 (13) \\
$\epsilon$ Vir & 5082 & 2.85 & 1.86 & $+$0.12 & $+$0.12 \\
HR 5058 & 4790 & 2.67 & 1.97 & $-$0.12 $\pm$ 0.13 (109) & $-$0.13 $\pm$ 0.22 (11) \\
\object{HR 5802} & 5010 & 2.89 & 1.67 & $-$0.02 $\pm$ 0.06 (129) & $-$0.03 $\pm$ 0.06 (13) \\
\object{HR 7321} & 4810 & 2.48 & 1.70 & $-$0.19 $\pm$ 0.06 (125) & $-$0.17 $\pm$ 0.07 (13) \\
\object{HR 8115} & 4910 & 2.41 & 1.85 & $-$0.04 $\pm$ 0.07 (125) & $-$0.03 $\pm$ 0.09 (13) \\
\object{HR 8204} & 5250 & 1.53 & 2.49 & $-$0.09 $\pm$ 0.12 (105) & $-$0.13 $\pm$ 0.11 (10) \\
\object{HD 205011} & 4780 & 2.41 & 1.70 & $-$0.14 $\pm$ 0.09 (117) & $-$0.13 $\pm$ 0.10 (13) \\
\object{HR 8878} & 4370 & 1.91 & 1.61 & $-$0.67 $\pm$ 0.07 (131) & $-$0.63 $\pm$ 0.10 (13) \\
\noalign{\smallskip}
\hline
\end{tabular}
\end{table*}

We selected two normal giants, HR 440 and HR 1326, for comparison
purposes, from the Bright Star Catalogue (Hoffleit \& Jaschek \@
\cite{BSC}). Judging by the color indices, their effective
temperatures should be similar to those of the barium giants. The
third normal star, HR 4932, is $\epsilon$ Vir, the same standard
star used by Zacs (\cite{Z94}) and one of the most extensively
studied giants in literature (Cayrel de Strobel et al.\@
\cite{CS2001}). We also adopted $\epsilon$ Vir as the standard
star for a differential analysis.

\begin{figure}
\centering
\includegraphics[width=7cm,height=6cm]{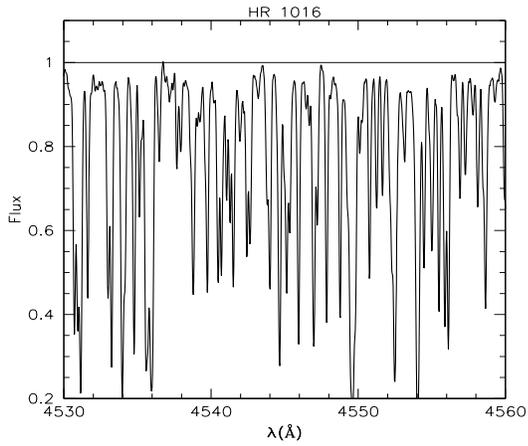}
\caption{An example of the normalization of the continuum in the bluest wavelength end used in this work.} \label{fig:con1}
\end{figure}

\begin{figure}
\centering
\includegraphics[width=7cm,height=6cm]{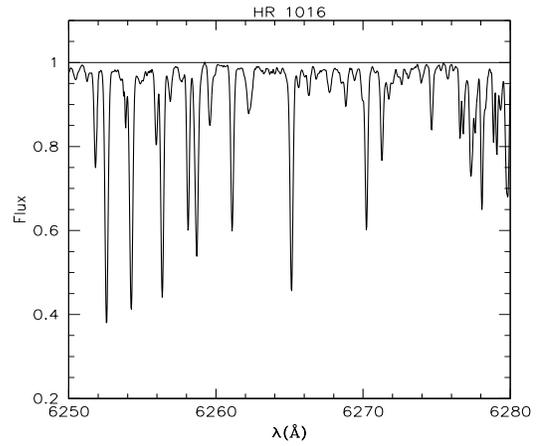}
\caption{A second example of the normalization of the continuum in a redder wavelength.} \label{fig:con2}
\end{figure}

The analysis was conducted using equivalent widths (EWs
hereafter). The EWs were measured in previously normalized sections of the spectra.
These sections were chosen to avoid spectral regions highly contaminated by telluric lines
and to avoid the wings of very strong lines that lower the continuum level. The pseudo-continuum level was carefully determined by identifying regions apparently free of spectral lines with the help of a high resolution solar spectrum atlas (Kurucz et al.\@ \cite{Kur84}). The continuum level was fitted with a low order Legendre polynomial crossing these regions. The crowding of the lines on the bluest end of the spectra makes this task more challenging than in the red end of the spectra. Nevertheless, we were able to find satisfactory fittings as exemplified in Figs. \ref{fig:con1} and \ref{fig:con2} for HR 1016.


For most of the elements we only used lines with EWs
smaller than 150 m\AA. In this case the EWs were determined by
fitting Gaussian profiles to the observed ones with
IRAF\footnote{IRAF is distributed by the National Optical
Astronomy Observatory, which is operated by the Association of
Universities for Research in Astronomy, Inc., under cooperative
agreement with the National Science Foundation.}.

\begin{figure}
\centering
\includegraphics[width=7cm,height=6cm]{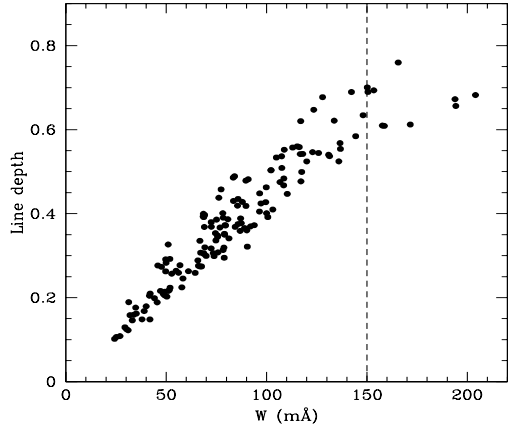}
\caption{Line depth vs.\@ EW for the Fe I lines of $\epsilon$ Vir.
This plot was used to determine the EW, 150 m\AA, where saturation
effects become important and a Gaussian profile is no longer the
best approximation of the line profile.} \label{fig:prof}
\end{figure}

This EW value was determined to be the limit of saturation of the curve of
growth. In this case the line growth is dominated by the Doppler
broadening, which determines a Gaussian profile and leads to
observed Gaussian profiles after convolution with the Gaussian
instrumental broadening profile. The region of linear growth can
be inferred in a plot of line depth vs.\@ EW. We have empirically
estimated the saturation limit by plotting the line depth vs.\@ EW
for several lines of the same chemical species. This was first
done for the Fe I lines in $\epsilon$ Vir, as shown in Figure
\ref{fig:prof}. A significant
departure from a linear growth happens only for EWs larger than
$\approx$ 150 m\AA. Similar plots were made for all stars and this
same limit proved reliable in all cases.

\begin{figure}
\centering
\includegraphics[width=7cm,height=6cm]{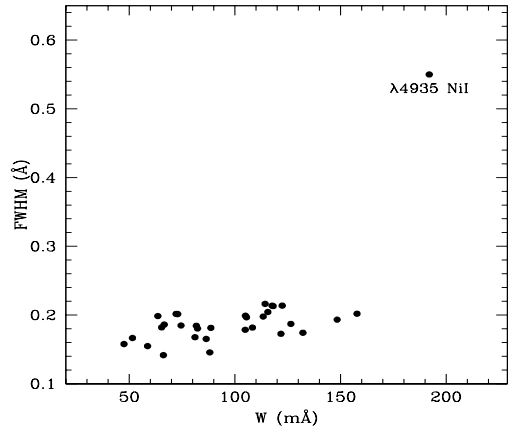}
\caption{Plot of FWHM vs.\@ EW for the lines of HR 5058 in the
$\lambda$$\lambda$4880-5165 spectral range. This plot was done for
all stars and used along with the plot of line depth vs. EW to
identify anomalous lines such as the NiI $\lambda$ 4935 line seen
in this plot. Lines departing from the expected trend were
excluded from the analysis.} \label{fig:fwhm}
\end{figure}

Lines stronger than 150 m\AA\@ were used only for elements with
only a few lines available throughout the spectra, namely Na, Mg,
Al, Cu, Zn, Sr, Y, Zr, Ba, La, Ce, and Nd. In this case the EWs
were determined by fitting Voigt profiles to the observed ones.
Slightly blended lines were measured using the deblending
capabilities of IRAF. The EWs of all the measured lines are listed
in the appendix (Tab. \ref{tab:le} and \ref{tab:le2}).

The plots of line depth vs.\@ EW were also used to test the
reliability of the measurements, along with plots of the full
width half maximum (FWHM) vs.\@ EW. For the latter we expect the
FWHM values to be distributed near the expected value for the
spectral resolving power, and to slowly increase with increasing
EW due to its progressive departure from a purely Gaussian
profile. Any line having, in any of these plots, a behavior
differing from the general expected trend was excluded from the
analysis. Figure \ref{fig:fwhm} shows an example of the FWHM vs.
EW plot for the star HR 5058, where the NiI $\lambda$ 4935 line is
clearly seen to not be well-fitted by a Gaussian profile, this
line then being excluded from the analysis of this star.

%____________________________ Atmospheric parameters ___________________________
%--- 3 ---

%
\begin{figure}
\centering
\includegraphics[width=7cm,height=6cm]{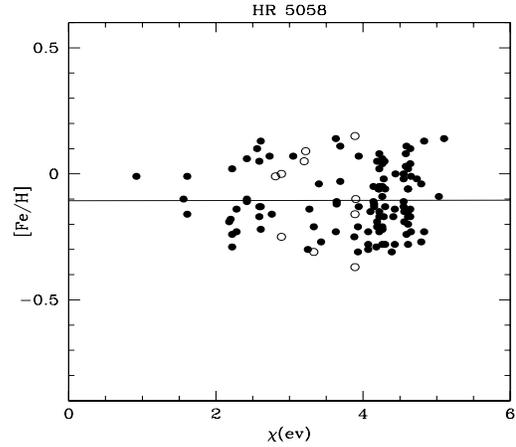}
\caption{Iron abundance of both Fe I (full circles) and Fe II
lines (open circles) vs.\@ the line excitation potential for the
star HR 5058. The solid line is a linear fit to the Fe I lines
indicating that the excitation equilibrium was fulfilled. The
ionization equilibrium was also obtained by setting the Fe I and
Fe II abundances to be equal, determining the surface gravity.}
\label{fig:fechi}
\end{figure}
\begin{figure}
\centering
\includegraphics[width=7cm,height=6cm]{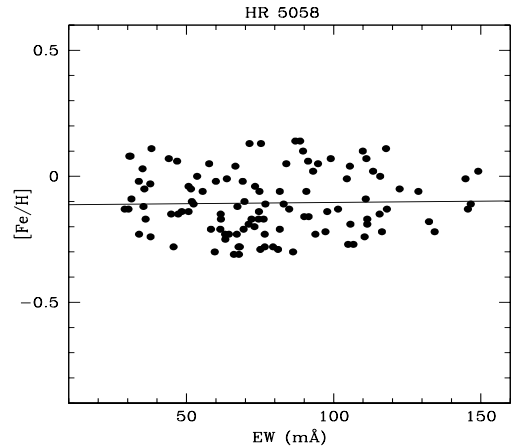}
\caption{Iron abundance vs.\@ EW for the Fe I lines of HR 5058.
This plot was used to determine the microturbulence velocity by
requiring a null correlation between [Fe/H] and the EWs.}
\label{fig:fele}
\end{figure}
\section{Atmospheric parameters}

\subsection{The differential analysis}

We conducted a model atmosphere analysis using the NMARCS grid
(Plez et al.\@ \cite{PBN92}, Edvardsson et al.\@ \cite{EDV1993}) 
originally developed by Gustafsson et al.\@ (\cite{G75}) and Bell et 
al.\@ (\cite{Be76}). The models assume a plane
parallel geometry,  local thermodynamic equilibrium (LTE) and
radiative equilibrium. We conducted a differential analysis using
$\epsilon$ Vir (HD 113226) as a standard star. $\epsilon$ Vir has
a solar like abundance pattern and is the same standard star used
in the analysis by Zacs (\cite{Z94}).

Within a differential analysis a robust scale for comparison of 
stellar parameters is established for similar stars. As will be seen 
in the following discussion, our sample defines a rather small range 
in the atmospheric parameters space, a fact that fully supports our 
approach. In this sense, we can safely disregard uncertainties due 
to the choice of model atmospheres, convection, inhomogeneities and 
NLTE since we expect these effects to be similar among the sample 
stars.

The effective temperature (T$_{\rm eff}$) for $\epsilon$ Vir was
calculated using the colors (R$-$I) (Hoffleit \& Jaschek \@
\cite{BSC}) and (V$-$K) (Johnson et al.\@ \cite{J66}) with the
metallicity independent relations by McWilliam (\cite{M90}), for
(R$-$I) and (V$-$K), and by Blackwell \& Lynas-Gray (\cite{B98}),
for (V$-$K). A mean T$_{\rm eff}$ was then calculated for the
(V-K) index, using the two calibrations, and this value was
averaged with the T$_{\rm eff}$ given by the (R-I) index. The
temperature thus obtained was T$_{\rm eff}$ = 5082 K. In a differential 
analysis the exact values of the parameters are of
much reduced importance, once the scales are homogeneous and
internally consistent.

With the V$_T$ magnitude from Tycho (ESA \cite{ESA}) and the
bolometric correction from Landolt \& B\"ornstein (\cite{LB82}) we
calculated the luminosity of $\epsilon$ Vir, log (L/L$_{\odot}$ )
= 1.83. The star was then placed in the HR diagram with
evolutionary tracks from Schaller et al.\@ (\cite{Sc92}) and
Schaerer et al.\@ (\cite{Sc93}). Assuming a solar metallicity a
mass estimate was then interpolated, M = 2.80 M$_{\odot}$.
Finally, using the well known equation, log g = log g$_{\odot}$  +
log (M/M$_{\odot}$) + 4 log (Tef/Tef$_{\odot}$) $-$ log
(L/L$_{\odot}$), we calculated its surface gravity, log g = 2.83.

The microturbulence velocity ($\xi$) was then determined
spectroscopically, by requiring the abundance obtained from Fe I
lines to have a null correlation with the EW. In this calculation
we firstly adopted laboratory \emph{gf}s for 18 Fe I lines given
by Blackwell et al.\@ (\cite{Bl95}) and Holweger et al.\@
(\cite{H95}). The microturbulence velocity thus obtained was $\xi$
= 1.86 km/s.

By calculating the atmospheric parameters we determine a unique
value for the metallicity. Using the above parameters a value of
[Fe/H] = +0.12 $\pm$ 0.08 was found. Since this value is different
from the one assumed to calculate the stellar mass, we
recalculated the parameters by adopting this new value for the
metallicity. The new parameters thus calculated are T$_{\rm eff}$
= 5082 K, log g = 2.85 dex, $\xi$ = 1.86 km/s, [Fe/H] = +0.12
$\pm$ 0.08, log(L/L$_{\odot}$)  = 1.83 and  M = 2.89 M$_{\odot}$.

From the quoted uncertainties in the calibrations used to derive the 
T$_{\rm eff}$, we estimate its 1$\sigma$ uncertainty to be 50K. The 
uncertainty in log g is estimated to be $\pm$ 0.13 dex. This value is 
obtained by the propagation of the uncertainties related to the quantities 
used in its calculation; $\pm$ 0.87 mas for $\pi$, $\pm$ 0.01 mag for the 
BC, $\pm$ 0.002 mag for V$_T$, $\pm$ 50 K for T$_{\rm eff}$, $\pm$ 0.2 
M$_{\odot}$ for the mass, and assuming negligible uncertainties in the 
solar parameters. Note that the mass uncertainty is only related to the 
error bars in log(L/L$_{\odot}$) and T$_{\rm eff}$, and does not take into 
account possible uncertainties inherent to the adopted tracks. Finally, 
the uncertainty in $\xi$ is estimated to be $\pm$ 0.08 km/s, given by 
the uncertainty in the slope of the correlation of [Fe/H] against EW.

This set of parameters for $\epsilon$ Vir is in good agreement with those previously
determined in the literature, as can be noted by comparison with the values listed in
 the catalogue by Cayrel de Strobel et al.\@ (\cite{CS2001}). In particular
we note the parameters obtained by McWilliam (\cite{M90}),
and listed in Table \ref{tab:1}, T$_{\rm eff}$ = 5060 K, log g = 2.97,
[Fe/H] = +0.15, and $\xi$ = 1.90 km/s.

Using the parameters we determined above and EWs from the spectrum of $\epsilon$
Vir, we derived a new set of $gf$s, for the whole set of lines
employed in this work, by requiring its abundance pattern to be
solar (where we adopt the solar abundances from Anders \& Grevesse \cite{AG89}). The set of $gfs$ thus calculated was used to derive the
atmospheric parameters and abundances for the other stars. The
$gf$s are listed in the appendix (Tab. \ref{tab:le} and
\ref{tab:le2}) along with the EWs.

\subsection{Atmospheric parameters for the sample stars}

For the other sample stars, T$_{\rm eff}$ was calculated by
requiring a null correlation of the iron abundance as given by the
Fe I lines with the excitation potential ($\chi$), thus fulfilling
the excitation equilibrium. The surface gravity is found by
requiring both Fe I and Fe II lines to have the same mean
abundance, thus fulfilling the ionization equilibrium. The
microturbulence velocity is found by requiring the Fe I abundance
to have a null correlation with the equivalent widths. By
simultaneously constraining these parameters we also determine the
metallicity, [Fe/H]. The parameters thus obtained are listed in
Table \ref{tab:par}. It can be seen that the stars define a narrow
range of T$_{\rm eff}$, log g and [Fe/H].

Figures \ref{fig:fechi} and \ref{fig:fele} show examples of the
null correlation obtained between the iron abundance and the
excitation potential or the EWs, respectively. In Fig.\@
\ref{fig:fechi} we also show the Fe II lines to exemplify the
ionization equilibrium.

\subsection{Uncertainties of the atmospheric parameters}

The uncertainties of the atmospheric parameters were calculated
for a representative star, HR 2392, which has atmospheric
parameters lying close to the center of the range defined by the
whole sample.

When determining the parameters, namely T$_{\rm eff}$ and $\xi$,
from the Fe I lines, we search for a linear fit where the angular
coefficient is statistically null. The uncertainty of this
coefficient sets the uncertainty of the T$_{\rm eff}$ and $\xi$
determinations. To find the 1$\sigma$ uncertainty of these
parameters we change their values, respectively, in the Fe I
abundance vs.\@ line excitation potential, and the Fe I abundance
vs.\@ EW diagrams, until the angular coefficients of the linear
fit match their own uncertainty.

\begin{table}
\caption{Uncertainties of the atmospheric parameters and of the equivalent widths.}
\label{tab:uncer}
\centering
\begin{tabular}{c c c c}
\noalign{\smallskip}
\hline\hline
\noalign{\smallskip}
Parameter & $\sigma$ & Range of EW & $\sigma$ \\
\hline
T$_{\rm eff}$ (K) & $\pm$ 50 & EW $<$ 150 m\AA & $\pm$ 3 m\AA  \\
log g (dex) & $\pm$ 0.35 & 150 m\AA\@ $<$ EW $<$ 400 m\AA & $\pm$ 10 m\AA  \\
$\xi$ (km/s) & $\pm$ 0.06 & EW $>$ 400 m\AA & $\pm$ 20 m\AA \\
\noalign{\smallskip}
\hline
\end{tabular}
\end{table}

In order to find the 1$\sigma$ uncertainty of the surface gravity
a different approach is required. The mean Fe abundances as given
from the Fe I lines and from the Fe II lines have, in general,
different standard deviations. The gravity is then changed until
the difference between the mean abundances from Fe I and Fe II
equals the larger of the standard deviations. This change in the
gravity is considered to be the 1$\sigma$ uncertainty in log g.
The uncertainties thus calculated are listed in Table
\ref{tab:uncer}.

\begin{table*}
\caption{Mean elemental abundances for the program stars, followed
by the standard deviation, when applicable, and the number of
lines on which the abundance is based.} \label{tab:abun}
\centering
\begin{tabular}{l c c c c c c}
\noalign{\smallskip}
\hline\hline
\noalign{\smallskip}
[X/Fe] & HR 440 & HR 649 & HR 1016 & HR 1326 & HR 2392 & HR 4608 \\
\hline
Na & $-$0.28 (02) & $-$0.09 (02) & $-$0.17 (02) & $-$0.31 (02) & $-$0.15 (02) & $-$0.26 (02) \\
Mg &   +0.02 $\pm$ 0.07 (03) & $-$0.05 $\pm$ 0.08 (04) & $-$0.07 $\pm$ 0.05 (04) & $-$0.07 $\pm$ 0.01 (03) &   +0.15 $\pm$ 0.19 (04) & $-$0.04 $\pm$ 0.05 (04) \\
Al &   +0.04 (02) & $-$0.09 (02) & $-$0.12 (02) &   +0.02 (02) &   +0.03 (02) &   +0.01 (02) \\
Si &   +0.01 $\pm$ 0.07 (15) & $-$0.10 $\pm$ 0.08 (15) & $-$0.12 $\pm$ 0.07 (14) & $-$0.09 $\pm$ 0.09 (10) &    0.00 $\pm$ 0.12 (08) & $-$0.07 $\pm$ 0.04 (10) \\
Ca & $-$0.09 $\pm$ 0.08 (12) & $-$0.06 $\pm$ 0.08 (11) & $-$0.07 $\pm$ 0.03 (09) & $-$0.15 $\pm$ 0.08 (11) & $-$0.07 $\pm$ 0.08 (10) & $-$0.03 $\pm$ 0.06 (11) \\
Sc & $-$0.03 $\pm$ 0.09 (07) & $-$0.08 $\pm$ 0.09 (07) & $-$0.13 $\pm$ 0.08 (07) & $-$0.18 $\pm$ 0.08 (07) & $-$0.03 $\pm$ 0.08 (07) & $-$0.02 $\pm$ 0.04 (07) \\
Ti &    0.00 $\pm$ 0.06 (30) & $-$0.02 $\pm$ 0.07 (34) & $-$0.08 $\pm$ 0.06 (35) & $-$0.02 $\pm$ 0.08 (34) & $-$0.02 $\pm$ 0.14 (30) &   +0.01 $\pm$ 0.08 (36) \\
V  & $-$0.08 $\pm$ 0.06 (11) & $-$0.06 $\pm$ 0.09 (11) & $-$0.16 $\pm$ 0.05 (11) & $-$0.05 $\pm$ 0.18 (10) & $-$0.03 $\pm$ 0.07 (10) & $-$0.11 $\pm$ 0.05 (09) \\
Cr & $-$0.10 $\pm$ 0.05 (25) & $-$0.07 $\pm$ 0.08 (27) & $-$0.09 $\pm$ 0.07 (24) & $-$0.11 $\pm$ 0.08 (22) & $-$0.09 $\pm$ 0.17 (16) & $-$0.10 $\pm$ 0.06 (22) \\
Mn & $-$0.18 $\pm$ 0.05 (06) & $-$0.20 $\pm$ 0.07 (06) & $-$0.20 $\pm$ 0.06 (06) & $-$0.12 $\pm$ 0.09 (04) & $-$0.09 $\pm$ 0.12 (05) & $-$0.24 $\pm$ 0.13 (06) \\
Co &   +0.02 $\pm$ 0.06 (09) & $-$0.02 $\pm$ 0.06 (09) & $-$0.09 $\pm$ 0.10 (10) & $-$0.12 $\pm$ 0.07 (10) & $-$0.01 $\pm$ 0.09 (08) &    0.00 $\pm$ 0.06 (09) \\
Ni & $-$0.09 $\pm$ 0.06 (26) & $-$0.16 $\pm$ 0.06 (28) & $-$0.16 $\pm$ 0.07 (27) & $-$0.06 $\pm$ 0.10 (25) & $-$0.13 $\pm$ 0.13 (24) & $-$0.11 $\pm$ 0.05 (24) \\
Cu & $-$0.06 $\pm$ 0.17 (03) & $-$0.17 $\pm$ 0.26 (03) & $-$0.09 (02) & $-$0.03 $\pm$ 0.10 (03) & $-$0.04 (02) & $-$0.01 $\pm$ 0.42 (03) \\
Zn &   +0.09 (01) &   +0.05 (01) &   +0.05 (01) & $-$0.10 (01) & $-$0.07 (01) &   +0.04 (01) \\
Sr & $-$0.16 (01) &   +0.16 (01) &   +0.12 (01) &   +0.03 (01) &   +1.24 (01) &   +0.60 (01) \\
Y  & $-$0.23 $\pm$ 0.09 (05) &   +0.01 $\pm$ 0.09 (06) & $-$0.11 $\pm$ 0.04 (06) & $-$0.06 $\pm$ 0.09 (06) &   +1.23 $\pm$ 0.36 (06) &   +0.44 $\pm$ 0.10 (06) \\
Zr & $-$0.21 $\pm$ 0.06 (03) & $-$0.04 $\pm$ 0.25 (03) & $-$0.08 $\pm$ 0.09 (03) & $-$0.07 $\pm$ 0.11 (03) &   +1.04 $\pm$ 0.46 (03) &   +0.60 (02) \\
Ba & $-$0.28 $\pm$ 0.13 (03) &   +0.08 $\pm$ 0.17 (03) &    0.00 $\pm$ 0.12 (03) & $-$0.27 $\pm$ 0.16 (03) &   +1.17 $\pm$ 0.44 (03) &   +0.54 $\pm$ 0.31 (03) \\
La & $-$0.14 $\pm$ 0.16 (04) &   +0.05 $\pm$ 0.20 (03) & $-$0.06 $\pm$ 0.14 (04) & $-$0.09 $\pm$ 0.20 (04) &   +1.52 $\pm$ 0.34 (04) &   +0.57 $\pm$ 0.08 (04) \\
Ce & $-$0.10 $\pm$ 0.06 (05) &   +0.09 $\pm$ 0.10 (05) &   +0.05 $\pm$ 0.10 (05) &   +0.06 $\pm$ 0.15 (05) &   +1.48 $\pm$ 0.55 (05) &   +0.68 $\pm$ 0.25 (05) \\
Nd & $-$0.08 (02) &   +0.06 (02) &   +0.02 (02) &   +0.06 (02) &   +0.91 (02) &   +0.58 (02) \\
Sm & $-$0.19 (01) & $-$0.13 (01) & $-$0.12 (01) & $-$0.06 (01) &   +0.78 (01) &   +0.20 (01) \\
Eu &    --    --  &    --    --  & $-$0.01 (01) &   +0.04 (01) &   +0.53 (01) &   +0.24 (01) \\
Gd & $-$0.32 (01) &    --    --  & $-$0.09 (01) & $-$0.17 (01) &   +0.06 (01)) &    --   --  \\
\noalign{\smallskip}
\hline
\end{tabular}
\end{table*}
\setcounter{table}{3}
\begin{table*}
\caption{continued.}
%\label{tab:abun2}
\centering
\begin{tabular}{l c c c c c c}
\noalign{\smallskip}
\hline\hline
\noalign{\smallskip}
[X/Fe] & HR 5058 & HR 5802 & HR 7321 & HR 8115 & HR 8204 & HR 8878 \\
\hline
Na & $-$0.10 (02) & $-$0.17 (02) & $-$0.28 (02) & $-$0.05 (02) & $-$0.21 (02) & $-$0.21 (2) \\
Mg & $-$0.01$\pm$0.16 (4) & $-$0.03$\pm$0.07 (04) & $-$0.05$\pm$0.06 (04) & $-$0.10$\pm$0.04 (04) &   +0.04$\pm$0.21 (04) &   +0.19$\pm$0.09 (03) \\
Al & +0.05 (02) & $-$0.03 (02) & $-$0.07 (02) & $-$0.07 (02) &   +0.33 (01) &   +0.21 (01) \\
Si & +0.03$\pm$0.13 & $-$0.07$\pm$0.05 (11) & $-$0.02$\pm$0.08 (15) & $-$0.05$\pm$0.09 (14) & $-$0.02$\pm$0.15 (11) &   +0.16$\pm$0.04 (12) \\
Ca & $-$0.10$\pm$0.14 (08) & $-$0.09$\pm$0.03 (10) & $-$0.10$\pm$0.06 (12) & $-$0.08$\pm$0.06 (11) & $-$0.05$\pm$0.11 (07) &   +0.10$\pm$0.05 (11) \\
Sc & $-$0.02$\pm$0.08 (07) & $-$0.07$\pm$0.05 (07) & $-$0.13$\pm$0.07 (07) & $-$0.18$\pm$0.09 (07) & $-$0.33$\pm$0.08 (05) &   +0.05$\pm$0.06 (07) \\
Ti & $-$0.02$\pm$0.13 (32) & $-$0.06$\pm$0.07 (35) & $-$0.06$\pm$0.06 (34) & $-$0.12$\pm$0.07 (36) &    0.00$\pm$0.18 (29) &   +0.34$\pm$0.13 (30) \\
V  & +0.08$\pm$0.11 (10) & $-$0.06$\pm$0.04 (10) & $-$0.15$\pm$0.02 (11) & $-$0.21$\pm$0.05 (11) & $-$0.13$\pm$0.12 (09) &   +0.23$\pm$0.17 (11) \\
Cr & +0.04$\pm$0.28 (21) & $-$0.03$\pm$0.09 (25) & $-$0.06$\pm$0.11 (26) &   +0.17$\pm$0.12 (27) & $-$0.11$\pm$0.18 (18) & $-$0.05$\pm$0.08 (23) \\
Mn & $-$0.13$\pm$0.09 (04) & $-$0.17$\pm$0.08 (06) & $-$0.18$\pm$0.04 (05) & $-$0.23$\pm$0.08 (06) & $-$0.16$\pm$0.32 (07) & $-$0.24$\pm$0.05 (06) \\
Co & 0.00$\pm$0.05 (08) & $-$0.08$\pm$0.06 (10) & $-$0.07$\pm$0.04 (09) & $-$0.12$\pm$0.08 (10) & $-$0.03$\pm$0.22 (08) &   +0.07$\pm$0.07 (07) \\
Ni & $-$0.08$\pm$0.14 (23) & $-$0.13$\pm$0.03 (21) & $-$0.12$\pm$0.05 (26) & $-$0.16$\pm$0.05 (22) & $-$0.20$\pm$0.10 (22) & $-$0.06$\pm$0.07 (24) \\
Cu & +0.55 (02) & $-$0.03$\pm$0.05 (03) &   +0.06$\pm$0.09 (03) &   +0.12$\pm$0.13 (03) & $-$0.03 (02) &   +0.19 (02) \\
Zn & $-$0.21 (01) & $-$0.04 (01) &    0.00 (01) & $-$0.03 (01) &   +0.18 (01) &   +0.14 (01) \\
Sr &  +1.38 (01) & +0.70 (01) &   +0.51 (01) &   +0.49 (01) &   +2.21 (01) &   +0.08 (01) \\
Y  & +0.99$\pm$0.33 (06) &  +0.50$\pm$0.11 (06) &   +0.29$\pm$0.12 (06) &   +0.37$\pm$0.10 (06) &   +1.66$\pm$0.45 (05) & $-$0.02$\pm$0.13 (05) \\
Zr &  +0.83$\pm$0.36 (03) & +0.48$\pm$0.04 (03) &   +0.27$\pm$0.04 (03) &   +0.22$\pm$0.04 (03) &   +1.00$\pm$0.60 (03) & $-$0.07$\pm$0.15 (03) \\
Ba & +0.93$\pm$0.36 (03) &  +0.16$\pm$0.17 (03) &   +0.31$\pm$0.39 (03) &   +0.31$\pm$0.26 (03) &   +1.08$\pm$0.39 (03) & $-$0.38$\pm$0.14 (03) \\
La &  +1.28$\pm$0.43 (04) & +0.27$\pm$0.10 (04) &   +0.23$\pm$0.14 (04) &   +0.19$\pm$0.18 (04) &   +1.32$\pm$0.57 (04) & $-$0.12$\pm$0.07 (04) \\
Ce & +1.05$\pm$0.53 (05) &  +0.20$\pm$0.10 (04) &   +0.32$\pm$0.17 (05) &   +0.23$\pm$0.16 (05) &   +1.48$\pm$0.71 (05) &   +0.05$\pm$0.20 (05) \\
Nd &  +0.78 (02) & +0.12 (02) &   +0.15 (02) &   +0.11 (02) &   +1.01 (02) &   +0.10 (02) \\
Sm &  +0.56 (01) &  0.00 (01) &   +0.01 (01) & $-$0.11 (01) &   +0.16 (01) & $-$0.07 (01) \\
Eu &  +0.49 (01) &  0.00 (01) &     --   --  & $-$0.03 (01) &   +0.02 (01) &   +0.24 (01) \\
Gd &  $-$0.09 (01) &  --   --  & $-$0.24 (01) & $-$0.23 (01) &    --    --  & $-$0.29 (01) \\
\noalign{\smallskip}
\hline
\end{tabular}
\end{table*}

We also estimated the uncertainty of the measurements of the EWs
using two stars with very similar atmospheric parameters, HR 649
and HR 1016. For the lines measured with Gaussian profiles, under
the hypothesis that the two stars should have the same set of
atmospheric parameters, and thus EWs, any difference on the EWs
are due only to errors on the measurements. It is clear that such
a hypothesis tends to overestimate the uncertainty since actual
differences between the two sets of EWs are expected. However, the
uncertainty thus calculated is only $\pm$ 3 m\AA, which emphasizes
the quality of our data and the good internal consistency of the
measurements.

For the stronger lines measured with Voigt profiles we adopted a
different approach, clearly needed since most of these lines are
due to s-process elements, and thus expected to have different
intensities in the chosen stars. In this case, the uncertainty was
estimated by fitting each line a few times, at each time slightly
changing the limits in wavelength for the profile fitting, keeping
control of the FWHM of the line Gaussian core. In this sense we
can verify the robustness of the fitting routine and the
sensibility of the measurements on the adopted wavelength limits
in a statistically significant way. In this case the uncertainties
found were $\pm$ 10m \AA\@ for 150 m\AA\@ $<$ EW $<$ 400 m\AA\@
and $\pm$ 20 m\AA\@ for EW $>$ 400 m\AA.

%____________________________ Abundances ___________________________________________
%--- 4 ---

%
\section{Abundances, masses and ages}

The abundances were calculated based on the EWs and the $gf$s
previously derived. For the elements Mg, Sc, V, Mn, Co and Cu we
considered the hyperfine structure following Steffen (\cite{S85}).
The abundances are listed in Table \ref{tab:abun}. We discuss
below the uncertainties of the abundances and then the abundance pattern
of each star in comparison with previous analyses available in the literature.
In the figures \ref{fig:ab1} to \ref{fig:205011}, the error bar on
the abundances is always the larger value between the standard
deviation of the [X/Fe] abundance ratios given by the stellar
lines (listed in Table \ref{tab:abun}), and the theoretically calculated
uncertainties for each element, based on the
atmospheric parameter and EW uncertainties derived in section
\ref{sec:sigma}. These theoretical error calculations are listed
in Table \ref{tab:sigma}.

\subsection{Uncertainties of the abundances}\label{sec:sigma}

The abundances are subject to uncertainties coming from the
determination of the atmospheric parameters. In order to estimate
these uncertainties we change each atmospheric parameter by its
own error, keeping the other ones with the original adopted
values, and recalculate the abundances. In this way we measure the
effect of each parameter uncertainty in the abundances. We also
estimated the uncertainty in the abundances coming from the errors
in the measurement of the EWs. These effects are listed in Table
\ref{tab:sigma}. The calculations were done, again, for the star
HR 2392. Assuming that the effects of the uncertainties of the
parameters are independent, we can estimate a lower bound of the
total uncertainty with Eq. \ref{eq:sigma}. The total compounded
uncertainty is also listed in Table \ref{tab:sigma}.

\begin{equation}
\sigma_{total} = \sqrt{(\sigma_{\rm{Teff}})^2 + (\sigma_{log g})^2 + (\sigma_{\xi})^2 + (\sigma_{[Fe/H]})^2 + (\sigma_{W})^2}
\label{eq:sigma}
\end{equation}

In general each error source affects the abundances by less than
0.1 dex, except for the uncertainty introduced by log g in the
abundances derived from lines of singly ionized species. For this
reason, the uncertainty of abundances derived solely from
neutral species are usually smaller. The mean value of the
theoretically estimated $\sigma_{total}$ is close to 0.12 dex, and
does not exceed 0.20 dex.

However, the abundance uncertainties, as estimated by the
dispersion of the mean of individual line abundances, of the stars
with the largest excesses of s-process elements, sometimes
appreciably surpass the values of Table \ref{tab:sigma}. This is
the case of HR 2392, 5058 and 8204, with the elements Y, Zr, La
and Ce having dispersions of the mean reaching $\approx$0.6 dex.
This clearly reflects enhanced errors in measuring the very strong
lines of these elements, leading us to the conclusion that at
least for a few cases the abundance uncertainties given in Table
\ref{tab:sigma} may be underestimated. The larger uncertainties in
the abundances of these elements in these stars, notwithstanding,
do not affect appreciably any of the conclusions of our analysis,
specially the classification of these objects as mild or classical
barium stars.

\subsection{Abundances}
\subsection*{HR 440}

HR 440 is one of the selected normal giants. Its metallicity is
lower than solar, [Fe/H] = $-$0.34 dex. Its abundance pattern is
shown in Fig. \ref{fig:ab1}. We note it to be almost solar, the
only significant difference is a 2$\sigma$ level deficiency for
Na. We also note indications of small deficiencies of Mn, Y, Zr,
Ba and Gd that are not significant within 2$\sigma$.

%
%\begin{figure}
%\centering
%\includegraphics[width=8cm]{440.eps}
%\caption{The abundance pattern for HR 440.}
%\label{fig:440}
%\end{figure}

%
\subsection*{HR 649}

HR 649 has a metallicity slightly lower than solar, [Fe/H] =
$-$0.14 dex, and is considered in the literature to be a mild
barium star. However, its abundance pattern, shown in Fig.\@
\ref{fig:ab1}, is solar except for a 2$\sigma$-level deficiency of
Mn.

In the catalogue by Lu (\cite{L91}) it has an index Ba0.3. Barium
stars are classified according to their level of overabundance in
the range Ba0.1$-$5, in a scheme first proposed by Warner
(\cite{W65}) and further extended (Keenan \& Pitts \cite{KP80}),
where Ba5 indicates the largest overabundances. Stars with Ba
index less than Ba2 are the so-called mild barium stars.

This star is a confirmed binary system with a white dwarf
companion identified by B\"ohm-Vitense \& Johnson (\cite{BJ85}).
It is thus one of the stars thought to support the scenario to
explain the origin of the abundances in barium and mild barium
stars.

It was previously analyzed by Zacs (\cite{Z94}), who found Y, Ba,
and La to be overabundant and Zr and Nd to be normal. Except for
Zr and Nd his results are based on fewer lines than ours. We also
found Sr and Ce to be normal. Our results rely on higher S/N data
and more sophisticated methods to measure equivalent widths than
those of Zacs (\cite{Z94}). Thus we conclude HR 649 not to be a
mild barium star but a normal giant.

%
%\begin{figure}
%\centering
%\includegraphics[width=8cm]{649.eps}
%\caption{The abundance pattern for HR 649.}
%\label{fig:649}
%\end{figure}
%
%
\subsection*{HR 1016}

HR 1016 also has a metallicity slightly lower than solar, [Fe/H] =
$-$0.11 dex, and is also considered in the literature to be a mild
barium star. It was first analyzed by Pilachowski (\cite{P77}) who
found small s-process enhancements for Y, Zr, and Ce, and normal
abundances for Sr, Ba, La, Pr, and Nd. We found no record of a
radial velocity variability or white dwarf companion in the
literature. Our abundance pattern for this star is shown in Fig.\@
\ref{fig:ab1}. We found no indication of anomalous abundances of
any s-process element. Its only peculiarities are deficiencies at
the 2$\sigma$ level of Na and Mn (as in HR 649), and probably of
V. We thus conclude this to be a case similar to that of HR 649: a
normal giant misclassified as a mild barium star.

%
%\begin{figure}
%\centering
%\includegraphics[width=8cm]{1016.eps}
%\caption{The abundance pattern for HR 1016.}
%\label{fig:1016}
%\end{figure}
%
%
\subsection*{HR 1326}

HR 1326 is a supposedly normal giant included in the analysis. It
has a solar metallicity and an almost solar abundance pattern, as 
shown in Fig. \ref{fig:ab1}. There is no chemical peculiarity
except for an apparent deficiency of Na at a level higher than
2$\sigma$.

%
%\begin{figure}
%\centering
%\includegraphics[width=8cm]{1326.eps}
%\caption{The abundance pattern for HR 1326.}
%\label{fig:1326}
%\end{figure}
%
%
\subsection*{HR 2392}

\begin{figure*}
\centering
\includegraphics[width=15cm]{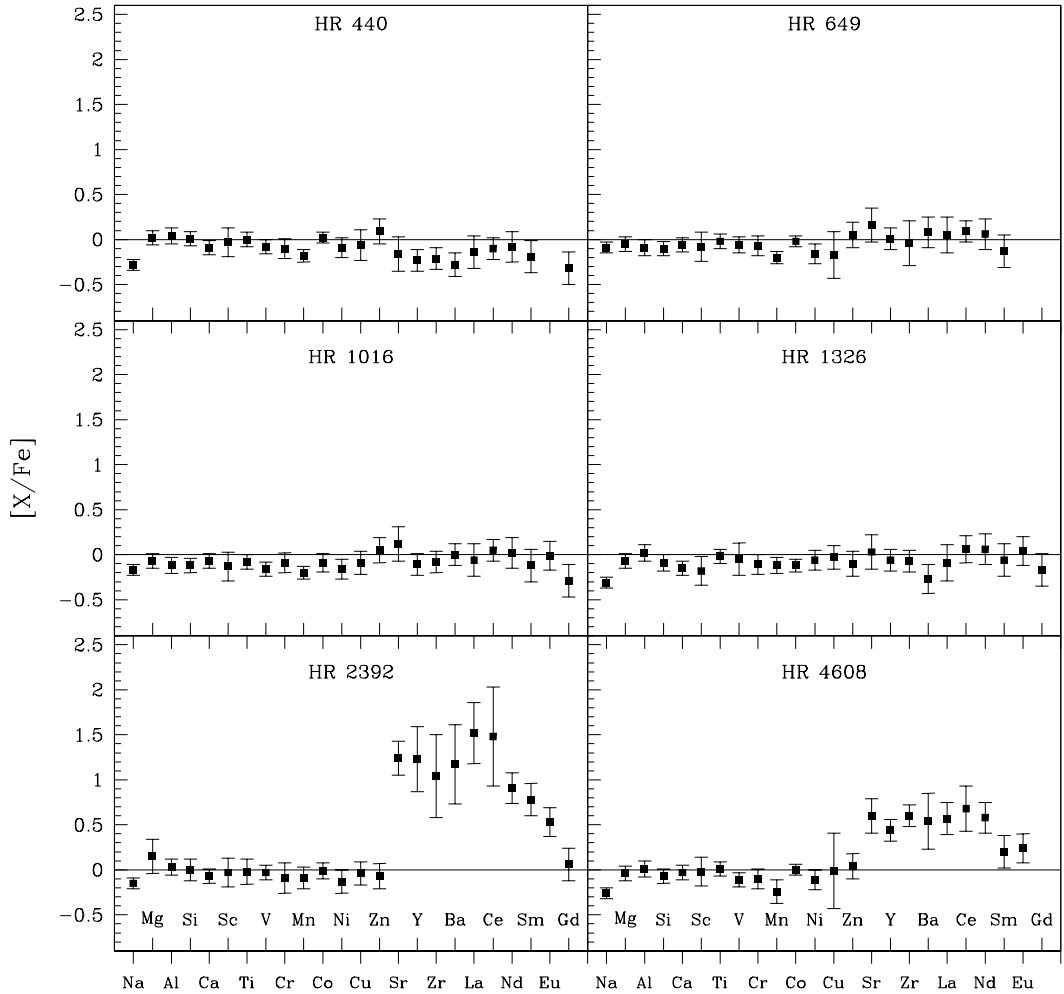}
\caption{The abundance pattern for HR 440, HR 649, HR 1016, HR 1326, HR 2392 and HR 4608.}
\label{fig:ab1}
\end{figure*}

HR 2392 (Fig.\@  \ref{fig:ab1}) is a classical barium star with [Fe/H] = $-$0.09. It was
 classified as such by Bidelman \& Keenan
(\cite{BK51}). According to Warner (\cite{W65}) it has a Ba3
index. McClure (\cite{Mc83}) found this star to have a variable
radial velocity and an orbital period of 457.7 $\pm$ 2.7 days.
However, the search for a white dwarf companion resulted in
negative detection with IUE (Dominy \& Lambert \cite{DL83}) and an
inconclusive detection with HST (B\"ohm-Vitense et al.\@
\cite{BV00})

The light elements all have solar levels except for Na, which
is deficient. In agreement with Zacs (\cite{Z94}) we found this
star to be overabundant in heavy elements, as shown in Fig.
\ref{fig:ab1}. The s-process elements have a mean overabundance
of +1.2 dex. The r-process dominated elements Sm and Eu are also
overabundant, whereas Gd shows a normal abundance.

%
%\begin{figure}
%\centering
%\includegraphics[width=8cm]{2392.eps}
%\caption{The abundance pattern for HR 2392.}
%\label{fig:2392}
%\end{figure}
%
%
\subsection*{HR 4608}

HR 4608 (Fig. \ref{fig:ab1}) is a mild barium star with [Fe/H] $-$0.35 dex  In the
catalogue by Lu (\cite{L91}) it has a Ba1 index. McClure
(\cite{Mc83}) has found no indication for radial velocity
variability, but B\"ohm-Vitense et al.\@ (\cite{BV00}) detected
UV-flux excesses that might indicate a white dwarf companion. Udry
et al.\@ (\cite{U98b}) derived a minimum period of P $>$ 4700 days
for this system. The light elements show a solar pattern, except
for a larger than 2$\sigma$ deficiency of Na. The s-process
elements show a mean overabundance of +0.6 dex. No r-process
overabundance can be established with statistical significance.

%
%\begin{figure}
%\centering
%\includegraphics[width=8cm]{4608.eps}
%\caption{The abundance pattern for HR 4608.}
%\label{fig:4608}
%\end{figure}
%

%
%\begin{figure}
%\centering
%\includegraphics[width=8cm]{5058.eps}
%\caption{The abundance pattern for HR 5058.}
%\label{fig:5058}
%\end{figure}
%

%
\subsection*{HR 5058}

HR 5058 (Fig.\@ \ref{fig:ab2}) is another classical barium star, first identified by
Bidelman \& Keenan (\cite{BK51}). Its metallicity is [Fe/H] =
$-$0.12, and Warner (\cite{W65}) classified this star as Ba3. HR
5058 is a member of a binary system with a white dwarf companion
identified by B\"ohm-Vitense et al.\@ (\cite{BV00}).

This star was previously analyzed by Luck \& Bond (\cite{LB91}).
Our results are in good agreement with theirs. The s-process
elements have a mean overabundance of +1.0 dex. The abundance
pattern shows a surprising overabundance of Cu. We can not compare the abundance
of Cu with previous results since ours seems to be the first Cu
abundance determination for this star, therefore we caution that
this intriguing result should await further confirmation. The
r-process elements Sm and Gd show a statistically significant
overabundance, $\sim$ 0.5 dex, whereas Gd seems to be normal,
similarly to what was found for HR 2392.

\subsection*{HR 5802}

\begin{figure*}
\centering
\includegraphics[width=15cm]{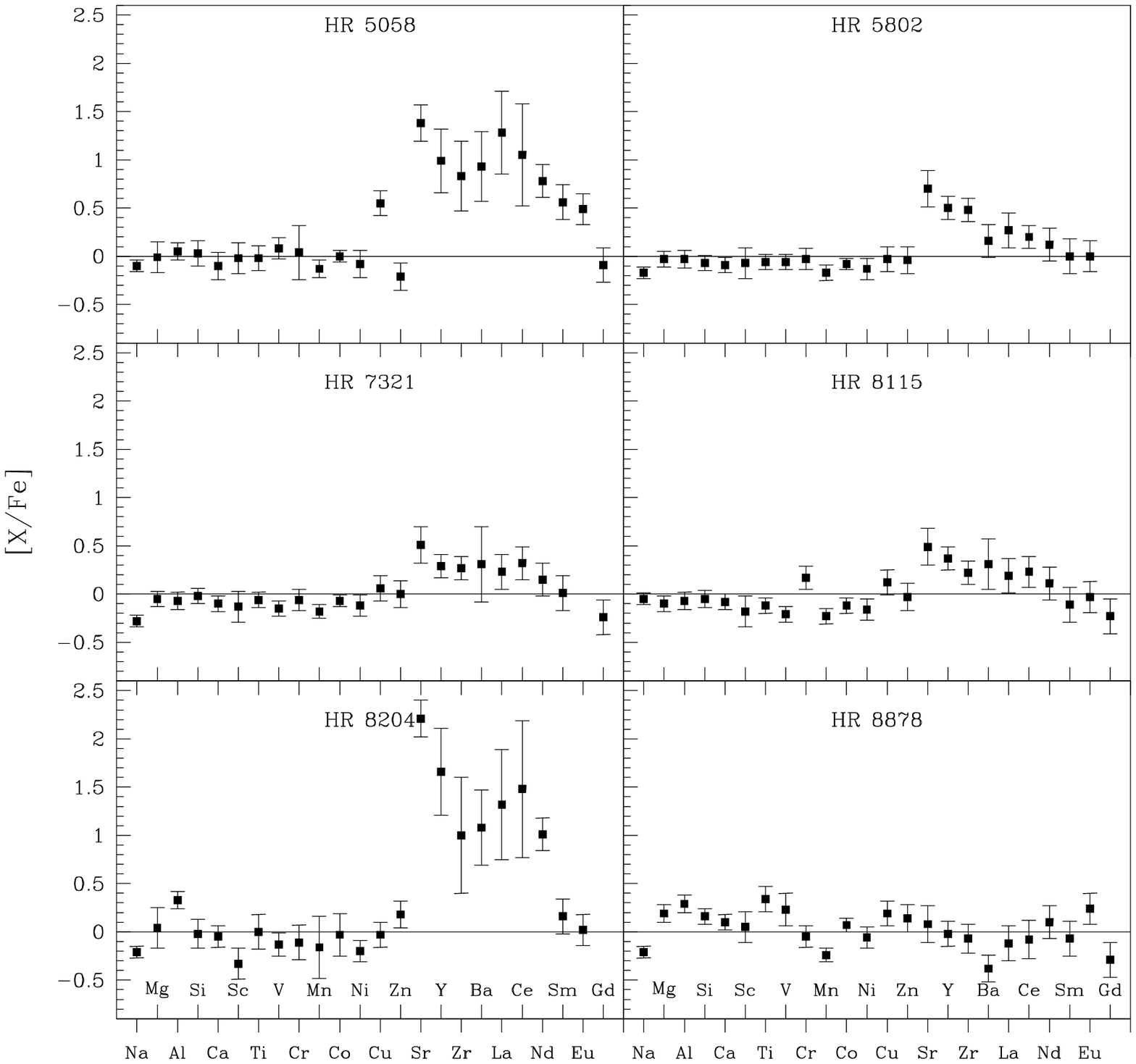}
\caption{The abundance pattern for HR 5058, HR 5802, HR 7321, HR 8115, HR 8204 and HR 8878.}
\label{fig:ab2}
\end{figure*}

HR 5802 is another mild barium star classified by Warner
(\cite{W65}) as Ba1.0. It has a solar metallicity. McClure
(\cite{Mc83}) argues this star to have a variable radial velocity.
Its abundance pattern is shown in Fig. \ref{fig:ab2} where we
note the light elements to follow a solar pattern except for a
larger than 2$\sigma$ deficiency of Na. In this star the lighter
s-process elements (Sr, Y and Zr) have a marked overabundance,
with a mean of +0.56 dex, while the heavier ones (Ba, La, Ce and
Nd) seem to be only slightly overabundant although statistically
they are normal.

Zacs (\cite{Z94}) analyzed this star and found results that agree
with ours within the uncertainties. There are only two exceptions,
Zr for which he found solar abundance and Ba for which he found
a larger excess.

%
%\begin{figure}
%\centering
%\includegraphics[width=8cm]{5802.eps}
%\caption{The abundance pattern for HR 5802.}
%\label{fig:5802}
%\end{figure}
%
%
\setcounter{table}{3}
\begin{table}
\caption{continued.}
%\label{tab:abun3}
\centering
\begin{tabular}{l c}
\noalign{\smallskip}
\hline\hline
\noalign{\smallskip}
[X/Fe] & HD 205011 \\
\hline
Na & $-$0.21 (02) \\
Mg & $-$0.08$\pm$0.04 (03) \\
Al & +0.01 (02) \\
Si & +0.05$\pm$0.09 (11) \\
Ca & $-$0.12$\pm$0.06 (11) \\
Sc & $-$0.15$\pm$0.10 (07) \\
Ti & $-$0.12$\pm$0.09 (33) \\
V & $-$0.18$\pm$0.04 (11) \\
Cr & $-$0.03$\pm$0.20 (25) \\
Mn & $-$0.21$\pm$0.09 (07) \\
Co & $-$0.12$\pm$0.09 (10) \\
Ni & $-$0.15$\pm$0.05 (25) \\
Cu & +0.12$\pm$0.14 (03) \\
Zn & $-$0.02 (01) \\
Sr & +0.89 (01) \\
Y & +0.90$\pm$0.32 (05) \\
Zr & +0.57$\pm$0.16 (03) \\
Ba & +0.66$\pm$0.34 (03) \\
La & +0.62$\pm$0.20 (04) \\
Ce & +0.63$\pm$0.33 (05) \\
Nd & +0.34 (02) \\
Sm & +0.09 (01) \\
Eu & +0.13 (01) \\
Gd & $-$0.13 (01) \\
\noalign{\smallskip}
\hline
\end{tabular}
\end{table}
\subsection*{HR 7321}

HR 7321 is another example of a mild barium star with metallicity
slightly lower than solar, [Fe/H] = $-$0.19 dex. Its abundance
pattern is shown in Fig. \ref{fig:ab2}. In the catalogue of Lu
(\cite{L91}) it has a Ba0.5 index. Most elements follow a solar
pattern except for the larger than 2$\sigma$ deficiencies of Na
and Mn. There is a good agreement between our abundances and those
by Zacs (\cite{Z94}), except for Zr which he founds to be
underabundant. As for HR 5802 the heavier s-process elements are
statistically normal while the lighter ones have only a slightly
larger than 2$\sigma$ enhancement with a mean value of +0.35 dex.

%
%\begin{figure}
%\centering
%\includegraphics[width=8cm]{7321.eps}
%\caption{The abundance pattern for HR 7321.}
%\label{fig:7321}
%\end{figure}
%

%
\begin{table}
\caption{The uncertainty of the abundances derived from the
uncertainties in the atmospheric parameters and EWs.}
\label{tab:sigma} \centering
\begin{tabular}{ccccccc}
\noalign{\smallskip}
\hline\hline
\noalign{\smallskip}
Element & $\sigma_{T_{\rm{eff}}}$  & $\sigma_{log g}$ & $\sigma_{\xi}$ & $\sigma_{[Fe/H]}$ & $\sigma_{W}$ & $\sigma_{total}$\\
\hline
Na & +0.03 & $-$0.02 & $-$0.02 & 0.00 & +0.04 & 0.06 \\
Mg & +0.04 & $-$0.05 & $-$0.02 & 0.00 & +0.04 & 0.08 \\
Al & +0.03 & $-$0.01 & $-$0.01 & $-$0.01 & +0.08 & 0.09 \\
Si & 0.00 & +0.06 & $-$0.01 & +0.01 & +0.05 & 0.08 \\
Ca & +0.04 & $-$0.03 & $-$0.03 & $-$0.01 & +0.05 & 0.08 \\
Sc & $-$0.01 & +0.15 & $-$0.01 & +0.03 & +0.04 & 0.16 \\
Ti & +0.05 & +0.02 & $-$0.03 & 0.00 & +0.05 & 0.08 \\
V & +0.07 & $-$0.01 & $-$0.01 & 0.00 & +0.04 & 0.08 \\
Cr & +0.04 & +0.03 & $-$0.02 & +0.07 & +0.06 & 0.11 \\
Mn & +0.05 & 0.00 & $-$0.02 & 0.00 & +0.04 & 0.07 \\
FeI & +0.03 & +0.02 & $-$0.02 & 0.00 & +0.05 & 0.06 \\
FeII & $-$0.04 & +0.17 & $-$0.03 & +0.03 & +0.06 & 0.19 \\
Co & +0.03 & +0.04 & 0.00 & +0.01 & +0.04 & 0.06 \\
Ni & +0.05 & +0.07 & 0.00 & +0.03 & +0.07 & 0.11 \\
Cu & +0.04 & +0.05 & $-$0.03 & +0.01 & +0.11 & 0.13 \\
Zn & $-$0.01 & +0.11 & $-$0.03 & +0.03 & +0.07 & 0.14 \\
Sr & +0.08 & $-$0.08 & $-$0.04 & +0.02 & +0.14 & 0.19 \\
Y & +0.01 & +0.09 & $-$0.03 & +0.04 & +0.07 & 0.12 \\
Zr & +0.03 & +0.08 & $-$0.06 & +0.01 & +0.06 & 0.12 \\
Ba & +0.01 & +0.01 & $-$0.01 & +0.05 & +0.03 & 0.06 \\
La & +0.02 & +0.12 & $-$0.06 & +0.03 & +0.12 & 0.18 \\
Ce & +0.01 & +0.09 & $-$0.04 & +0.03 & +0.07 & 0.12 \\
Nd & +0.02 & +0.15 & $-$0.04 & +0.03 & +0.07 & 0.17 \\
Sm & +0.02 & +0.15 & $-$0.04 & +0.03 & +0.08 & 0.18 \\
Eu & 0.00 & +0.15 & $-$0.02 & +0.03 & +0.05 & 0.16 \\
Gd & 0.00 & +0.15 & $-$0.01 & +0.02 & +0.09 & 0.18 \\
\noalign{\smallskip}
\hline
\end{tabular}
\end{table}

\subsection*{HR 8115}

HR 8115 (Fig.\@ \ref{fig:ab2}) is another example of a mild barium star and was
classified as Ba1 by Lu (\cite{L91}). Griffin \& Keenan
(\cite{GK92}) found the star to be a binary with a rather long
period, 18 years. Dominy \& Lambert (\cite{DL83}) found a UV flux
excess that probably indicates a white dwarf companion.

Its metallicity is solar. Up to the iron-peak, the only deviations
from a solar abundance pattern are deficiencies of V and Mn at the
2$\sigma$ level. As in HR 5802 and HR 7321, the heavier s-process
elements seem to be slightly enhanced but with low statistical
significance. The lighter s-process elements, Sr, Y and
Zr seem to be more clearly enhanced, particularly Sr. Abundances
from Ba to Ce could be normal, and from Nd to Gd are solar. Our
abundances are in relatively good agreement with the abundances
derived by Pilachowski (\cite{P77}) and Yushchenko et al.\@
(\cite{Y04}). We tentatively suggest this star to be a mild barium
star.

%
%\begin{figure}
%\centering
%\includegraphics[width=8cm]{8115.eps}
%\caption{The abundance pattern for HR 8115.}
%\label{fig:8115}
%\end{figure}
%
%
\subsection*{HR 8204}

HR 8204 (Fig.\@ \ref{fig:ab2}) is a classical barium star with [Fe/H] = $-$0.09 dex. It was
first recognized by Bidelman \& Keenan (\cite{BK51}) and was the
first barium system where a UV flux excess was identified
(B\"ohm-Vitense \cite{BV80}). It is classified by Warner
(\cite{W65}) as Ba2.

The light elements show a solar pattern, except by a larger than 2
$\sigma$ underabundance of Na and an overabundance of Al. Sc is
probably deficient, and all the s-process elements have
overabundances larger than +1.0 dex, with Sr reaching +2.2 dex;
the mean overabundance is +1.4 dex. This star was previously
analyzed by Zacs (\cite{Z94}) and our results are in agreement
with his within the uncertainties. The r-process elements Sm and
Eu are not overabundant.

%
%\begin{figure}
%\centering
%\includegraphics[width=8cm]{8204.eps}
%\caption{The abundance pattern for HR 8204.}
%\label{fig:8204}
%\end{figure}
%

%
\subsection*{HR 8878}

HR 8878 (Fig.\@ \ref{fig:ab2}) is the most metal deficient star in our analysis, [Fe/H] =
$-$0.67. Its abundance pattern is enriched in Ti, Mg and Al,
while Na is deficient. Mn is also deficient at the 2$\sigma$
level. These values are expected for normal metal deficient stars,
except for Na, which should track Al (Edvardsson et al.
\cite{EDV1993}, McWilliam \cite{M97}). We found no s-process
element to be overabundant, in contrast with Zacs (\cite{Z94}),
who found Ba to be enhanced, we found a deficiency of
this element at the 2$\sigma$ level, as expected for mildly metal
poor stars. Our abundances classify this object as a normal metal
poor giant star.

%
%\begin{figure}
%\centering
%\includegraphics[width=8cm]{8878.eps}
%\caption{The abundance pattern for HR 8878.}
%\label{fig:8878}
%\end{figure}
%
%
\begin{figure}
\centering
\includegraphics[width=7cm,height=6cm]{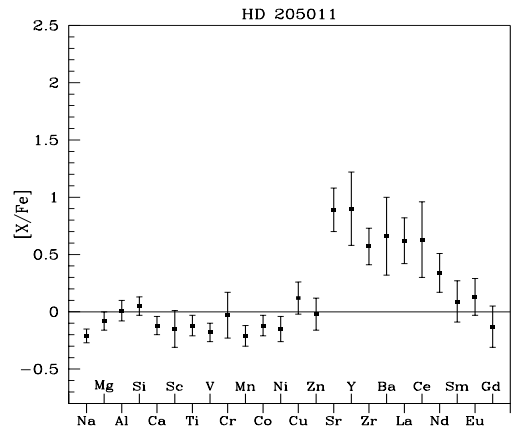}
\caption{The abundance pattern for HD 205011.}
\label{fig:205011}
\end{figure}
\begin{table*}
\caption{Visual magnitude, V$_T$, parallax, absolute magnitude
M$_{VT}$, bolometric correction BC, bolometric magnitudes and
luminosities of the sample stars. The last column refers to the
luminosity uncertainty due to error propagation of the involved
quantities.} \label{tab:lum} \centering
\begin{tabular}{c c c c c c c c c }
\noalign{\smallskip}
\hline\hline
Star & V$_T$ & $\pi$ (mas) & M$_{VT}$ & BC & M$_{BOL}$ & $\sigma_{M_{BOL}}$ & log (L$_{\star}$ / L$_{\odot}$) & $\sigma_{log (L_{\star} / L_{\odot})}$ \\
\noalign{\smallskip}
\hline
HR 440 & 4.05 & 22.15 & +0.78 & $-$0.34 & +0.44 & 0.04 & 1.75 & 0.02 \\
HR 649 & 4.47 & 9.01 & $-$0.76 & $-$0.22 & $-$0.98 & 0.21 & 2.32 & 0.09 \\
HR 1016 & 5.60 & 7.75 & +0.04 & $-$0.23 & $-$0.19 & 0.14 & 2.00 & 0.05 \\
HR 1326 & 3.98 & 27.85 & +1.21 & $-$0.39 & $-$0.81 & 0.03 & 1.60 & 0.01 \\
HR 2392 & 6.39 & 8.25 & +0.97 & $-$0.28 & +0.68 & 0.16 & 1.65 & 0.07 \\
HR 4608 & 4.22 & 19.08 & +0.63 & $-$0.29 & +0.34 & 0.06 & 1.79 & 0.03 \\
$\epsilon$ Vir & 2.95 & 31.90 & +0.47 & $-$0.23 & +0.24 & 0.04 & 1.83 & 0.02 \\
HR 5058 & 5.24 & 15.73 & +1.23 & $-$0.34 & +0.89 & 0.08 & 1.57 & 0.03 \\
HR 5802 & 5.36 & 13.89 & +1.07 & $-$0.26 & +0.82 & 0.09 & 1.60 & 0.04 \\
HR 7321 & 6.52 & 6.68 & +0.64 & $-$0.33 & +0.31 & 0.21 & 1.80 & 0.08 \\
HR 8115 & 3.31 & 21.62 & $-$0.01 & $-$0.29 & $-$0.31 & 0.03 & 2.05 & 0.01 \\
HR 8204 & 3.86 & 8.19 & $-$1.57 & $-$0.18 & $-$1.76 & 0.18 & 2.63 & 0.07 \\
HD 205011 & 6.54 & 6.31 & +0.54 & $-$0.34 & +0.20 & 0.15 & 1.85 & 0.06 \\
HR 8878 & 5.21 & 9.56 & +0.11 & $-$0.52 & $-$0.41 & 0.13 & 2.09 & 0.05 \\
\noalign{\smallskip}
\hline
\end{tabular}
\end{table*}

\subsection*{HD 205011}

HD 205011 (Fig.\@ \ref{fig:205011}) is another mild barium star analyzed by Zacs
(\cite{Z94}) and classified by Lu (\cite{L91}) as Ba2. Its
metallicity is slightly lower than solar, [Fe/H] = $-$0.14 dex.
McClure (\cite{Mc83}) found the star to have a variable radial
velocity. Our abundances are generally in good agreement with
those by Zacs (\cite{Z94}) except for Zr which he found to be
solar. Na, V and Mn seem to have a larger than 2$\sigma$
deficiency. The s-process elements have a mean overabundance of
+0.66 dex, and this star seems to be intermediate between mild and
classical barium stars. The r-process elements have normal
abundances.

\subsection{Masses and ages}

We also estimated the masses and ages of the sample stars by
placing them in the HR diagram with isochrones and theoretical
evolutionary tracks from the Geneva group (Charbonnel et al.\@
\cite{Ch93}; Schaerer et al.\@ \cite{Sc93}; and Schaller et al.\@
\cite{Sc92}). We made use of the previously derived T$_{\rm eff}$
and calculated the luminosities as follows.

With parallaxes and visual magnitudes, V$_{T}$, from the Hipparcos
catalogue (ESA \cite{ESA}), and bolometric corrections from
Landolt-B\"ornstein (\cite{LB82}), we calculated the absolute
bolometric magnitudes. Thus, adopting the solar bolometric
magnitude in the Tycho system, M$_{{Bol}_{\odot}}$ = 4.81, we
calculated the luminosities using $log(L_{\star} / L_{\odot}) =
-0.4(M_{{Bol}_{\star}} - M_{{Bol}_{\odot}})$. The magnitudes and
luminosities with respective uncertainties are listed in Table 
\ref{tab:lum}.

\begin{figure}
\centering
\includegraphics[width=7cm,height=6cm]{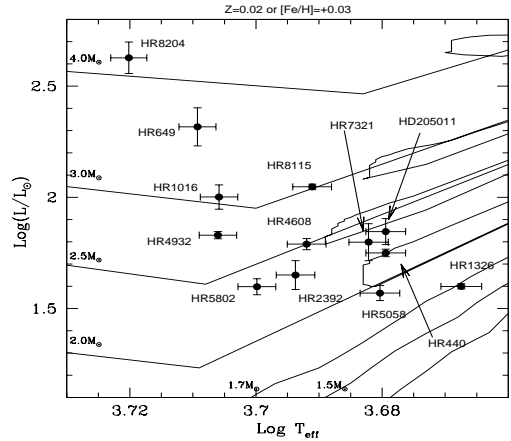}
\caption{The HR diagram with our stars and theoretical tracks from
Schaller et al.\@ (\cite{Sc92}) for [Fe/H] = +0.03. The most metal
deficient star of our sample, HR 8878, is not shown in this plot. The normal stars
are shown as open squares, the mild barium stars as circles and the barium
stars as triangles.}
\label{fig:hr}
\end{figure}
\begin{table}
\caption{Masses, ages and an evolutionary estimates of the surface
gravity of the sample stars.} \label{tab:mass} \centering
\begin{tabular}{c c c c c}
\noalign{\smallskip}
\hline\hline
Star & Mass (M$_{\odot}$) & Age (Gy) & log g & $\sigma_{log g}$ \\
\noalign{\smallskip}
\hline
HR 440 & 1.9 & 0.8 & 2.64 & 0.03 \\
HR 649 & 3.6 & 0.1 & 2.47 & 0.10 \\
HR 1016 & 3.0 & 0.2 & 2.70 & 0.07 \\
HR 1326 & 1.5 & 1.3 & 2.65 & 0.03 \\
HR 2392 & 2.3 & 0.5 & 2.87 & 0.08 \\
HR 4608 & 2.3 & 0.5 & 2.72 & 0.04 \\
$\epsilon$ Vir & 2.9 & 0.3 & 2.85 & 0.03 \\
HR 5058 & 1.9 & 0.9 & 2.82 & 0.05 \\
HR 5802 & 2.4 & 0.5 & 2.97 & 0.05 \\
HR 7321 & 2.2 & 0.5 & 2.66 & 0.10 \\
HR 8115 & 3.0 & 0.2 & 2.58 & 0.03 \\
HR 8204 & 4.2 & 1.0 & 2.27 & 0.09 \\
HD 205011 & 2.2 & 0.5 & 2.61 & 0.08 \\
HR 8878 & 1.0 & 6.2 & 1.87 & 0.07 \\
\noalign{\smallskip}
\hline
\end{tabular}
\end{table}

For all stars but HR 8878, the masses and ages were linearly
interpolated from the values obtained by placing the stars in
theoretical diagrams for two metallicities, [Fe/H] = $-$0.40 and
[Fe/H] = +0.00. The mass and age of HR 8878 were obtained by using
a third diagram with [Fe/H] = $-$0.72. The masses and ages thus
calculated are listed in Table \ref{tab:mass}. Figure \ref{fig:hr}
shows the region of the HR diagram where the sample stars are
located.

We also estimated the ``astrometric" log gs using the derived
masses of the stars, as done for $\epsilon$ Vir in section 3.1.
The log gs and its uncertainty, calculated in a straightforward
way by the error propagation of the involved quantities, are also
listed in Table \ref{tab:mass}. The agreement between the
ionization and astrometric values of log g is good within the
uncertainties, except for HR 8204, which has a significantly lower
spectroscopic log g. The barium and mild barium giants of our
sample define together a narrow range in stellar mass, 1.9 to 4.2
M$_{\odot}$, and age, 0.2 to 1.0 Gyr, indicating that they most
probably share similar evolutionary stages.

\section{Discussion}
\subsection{The mild barium stars}

With the first few orbital parameters for barium stars, McClure et
al.\@ (\cite{MFN80}) were able to confirm the binarity of most of
the strong barium stars (Ba$>$2) they analyzed. Based on that it
was possible to conclude that the strong barium stars were all
probable members of binary systems.

The same conclusion, however, was not possible in the case of the
mild barium stars. Most of the stars for which no radial velocity
variability was found belonged to the mild barium class. A firm
conclusion was not possible although they mostly seemed to be
members of a binary system (McClure \cite{Mc84}).

It was then suggested that mild barium stars could have wider
separations between the components (B\"ohm-Vitense et al.\@
\cite{BV84}), which would require observations spanning longer
durations to detect their binary nature. Some mild barium stars
were, however, found to be misclassified, as in Smith \& Lambert
(\cite{SL87}). Naturally this leads to the question of how many
misclassified objects there are in the mild barium star lists.

More recently, based on an extended sample of orbital elements
(Udry et al.\@ \cite{U98a}, \cite{U98b}), Jorissen et al.\@
(\cite{J98}) were able to confirm the binary status for 34 among
40 mild barium stars. Thus it was shown that the frequency of
detected binaries is compatible with all mild barium stars
belonging to binary systems. However the frequency of detected
binarity in strong barium stars is still larger, 35 out of 37.

Jorissen et al.\@ (\cite{J98}) have also shown that mild barium
stars are not restricted to long period systems. This argues
against the suggestion that the non-detection of binarity in some
mild barium systems is due solely to wider separations. In this
work we found the stars HR 649 and HR 1016 not to be mild barium
stars, in contrast to what was generally accepted in the
literature. Moreover, since HR 649 seems to have a white dwarf
companion (B\"ohm-Vitense \& Johnson \cite{BJ85}) the very fact it
exists shows that binarity is not a sufficient condition to form a
barium system.

Particularly we note that many works on general characteristics of
barium (and mild barium stars) such as kinematics (G\'omez et
al.\@ \cite{G97}) or distribution along the HR diagram (Bergeat \&
Knapik \cite{BK97}) adopted lists of stars that have never been
targets of an abundance analysis based on high resolution spectra.
Thus, many of the stars have only a tentative classification as
mild barium. If a significant number of tentative mild barium
stars are in fact misclassified, these studies may be suffering
from important biases.

The results by Jorissen et al.\@ (\cite{J98}) indicate that a
wider separation is not the parameter controlling the difference
in the abundances between mild and strong barium stars. They argue
that metallicity may be the main parameter. Thus, the increasing
level of overabundances seen in mild barium, strong barium and
population II CH-stars would correspond to a sequence of older,
and thus more metal-poor, populations. A discussion of this
suggestion is presented below but we can state in advance that
barium and mild barium stars do not seem to have different
metallicities.

Thus, there are still important open questions on the nature of
the mild barium systems. Moreover, there seems to be no work in
the literature on confirming the abundance peculiarities of a
large sample of mild barium stars. We are currently analyzing a
large sample of southern, tentative mild barium systems in order
to derive their barium abundances, their metallicities and
atmospheric parameters. These results and a discussion on this
subject will be presented in a forthcoming paper.

\subsection{What differentiates barium and mild barium stars?}

The s-process nucleosynthesis has long being recognized to be a
complex process. In order to explain the observations it was
divided into three components, the main, strong and weak
components. The weak component is thought to occur during He
burning in massive stars and to be responsible for isotopes from
Fe to Sr. The main and strong components are thought to occur in
AGB stars (Busso et al.\@ \cite{B99}). The strong component is
needed to build $^{208}$Pb, which has been shown to happen in
metal-poor AGBs (Travaglio et al.\@ \cite{T01}). The main
component, in low mass AGBs, is necessary to build s-elements from
Sr to Pb.

In low mass AGBs the neutrons flux seems to be mostly due to the
$^{13}$C($\alpha$,$n$)$^{16}$O reaction in a radiative layer
during the interval between thermal pulses. A marginal activation
of the $^{22}$Ne($\alpha$,$n$)$^{25}$Mg reaction in convective
conditions during the thermal pulses may also occur (Busso et
al.\@ \cite{B01}).

Barium (and mild barium) stars are thought to be the result of
mass transfer from s-process enriched AGB stars. Thus their
overabundances should follow the general pattern expected for
AGBs, as was shown elsewhere (Busso et al.\@ \cite{B95};
\cite{B01}). Following these results, we will here compare and
discuss the observed patterns in barium and mild barium stars, in
order to highlight some differences and similarities between these
groups of stars.

Following Luck \& Bond (\cite{LB91}), the [hs/ls] ratio has been
widely used as an indicator of s-process efficiency. [hs] stands
for the mean abundance of the "heavy" s-process elements of the Ba
peak, and [ls] is the same for the "light" s-process elements of
the Zr peak. These nuclei have neutron magic numbers and thus have
a low cross section against further neutron captures.

Various authors adopted different elements to calculate [hs] and
[ls]. Here we use Sr, Zr and Y to calculate [ls] and Ba, La, Ce
and Nd to calculate [hs]. We calculated this ratio for our stars
and also for stars analyzed in the literature with high
resolution, high signal to noise spectra (Boyarchuk et al.\@
\cite{B02}; Liang et al.\@ \cite{Li03}; Antipova et al.\@
\cite{A04}) whenever abundances for the same elements were
available.

\begin{figure}
\centering
\includegraphics[width=7cm,height=6cm]{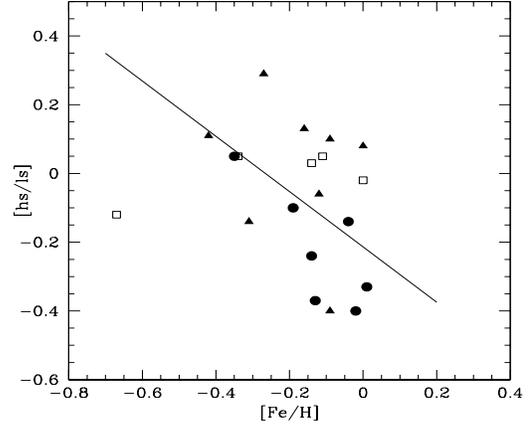}
\caption{The ratio [hs/ls] vs. [Fe/H] for our sample stars.
hs/Fe] is the mean abundance of Ba, La, Ce and Nd, [ls/Fe] is the
mean of Sr, Zr and Y. In this plot the normal stars are
represented as open squares, the mild barium stars as circles and
the barium stars as triangles. The solid line is a linear fit to
the peculiar giants only, barium and mild barium, excluding the
normal giants.} \label{fig:hsls}
\end{figure}

This ratio has been shown to be a (complex) function of the
neutron exposure (Busso et al.\@ \cite{B01}). Fig. \ref{fig:hsls}
shows the plot of [hs/ls] vs. [Fe/H]. In this figure the normal
giants are shown as open squares, the mild barium stars as circles
and the barium stars as triangles. The general trend observed in
several works is also clear here, there is an anticorrelation
between [hs/ls] and [Fe/H]. In the restricted metallicity range we
are considering here a direct correlation between [hs/ls] and the
neutron exposure is not possible (Busso et al.\@ \cite{B01}),
since the several theoretical tracks overlap. It is however
interesting to note that there is a general trend for the mild
barium stars to fall below the barium stars of same metallicity.

Although the scatter seems to be large, the standard deviation
from a linear fit (using only barium and mild barium stars) is
$\sigma$ = 0.20, a value that can be fully ascribed to
observational uncertainties. The few normal giants plotted show
little scatter ($\approx$0.08 dex) around [hs/ls] = 0.0 dex, as
would be expected for a solar scaled mixture.

\begin{figure}
\centering
\includegraphics[width=7cm,height=6cm]{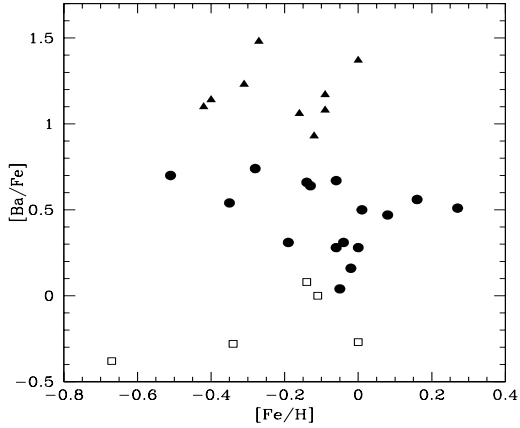}
\caption{The barium abundance, [Ba/Fe], vs. iron
abundance, [Fe/H]. The symbols have the same meaning as in Fig.19.
It is clear that there is no separation between barium and mild
barium stars with respect to the iron abundance.} \label{fig:bafe}
\end{figure}

An important fact that appears in Fig. \ref{fig:hsls} and is
clearer in Fig. \ref{fig:bafe} is that there seems to be no
significant difference in iron abundance between mild barium and
barium stars. The two groups have also the same metallicity range
of the normal disk giants considered here. The only separation
occurs in barium abundance as anticipated by the labels, normal,
mild barium and barium giants. Thus, at least with respect of
metal content, barium and mild barium stars seem to be members of
the same stellar population.

At lower metallicities higher overabundances of s-process elements
can be achieved (Busso et al.\@ \cite{B01}). This happens because
for a given neutron flux the neutron exposure increases with
decreasing number of iron group seed nuclei. In addition, in a
lower metallicity environment the abundance of the neutron
poisons, nuclei that do not take part in the s-process branches but
also capture neutrons, is also smaller. Thus it has been argued
that barium stars could be members of a more metal-poor population
than mild barium stars. The data presented here (Fig.
\ref{fig:hsls} and Fig \ref{fig:bafe}) do not support that
suggestion.

Having discarded the hypothesis of a more metal rich origin for
mild barium stars and that they are not restricted to long-period
systems (Jorissen et al.\@ \cite{J98}), other scenarios for their
formation should be found.

The essence behind the proposition of different metallicities to
explain the difference in overabundances lies in the different
neutron exposure that the materials would have been subjected to.
Even though we discarded this difference in the metallicities,
there could be another mechanism affecting the neutron exposure,
so that the material in mild barium stars could still be the
result of a smaller neutron exposure.

\begin{figure}
\centering
\includegraphics[width=7cm,height=6cm]{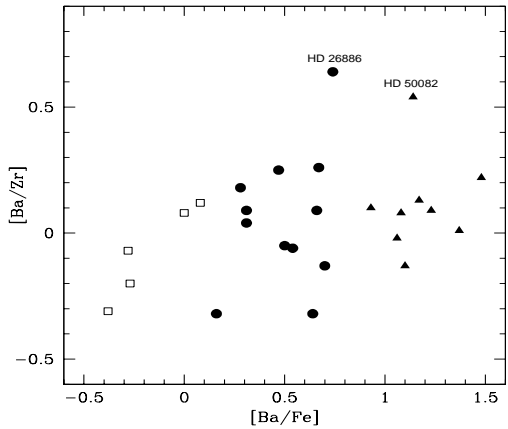}
\caption{Plot of [Ba/Zr] vs. [Ba/Fe] for the whole sample being
considered. The symbols are the same as in the previous plots.}
\label{fig:bazr}
\end{figure}

\begin{figure}
\centering
\includegraphics[width=7cm,height=6cm]{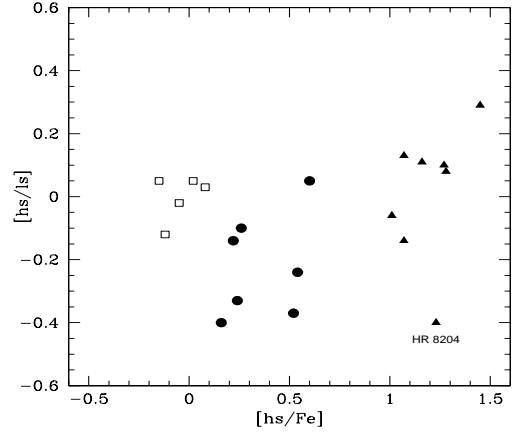}
\caption{Plot of [hs/ls] vs. [hs/Fe] for the whole sample being
considered. The symbols are the same as in the previous plots.}
\label{fig:hslshs}
\end{figure}

For a given metallicity, a lower neutron exposure should favor the
production of nuclei in the Zr peak. Increasing the neutron
exposures should favor the formation of the Ba peak nuclei. This
means that, for a given metallicity, the ratio of the Ba peak to
the Zr peak abundances ([Ba/Zr] or [hs/ls]) should increase with
increasing neutron exposure and hence with increasing Ba peak
abundances. One has to note, however, that for a given neutron
exposure a decreasing metallicity would produce the same effect.
Thus, we search for any significant sign of this possible effect
in Figs. \ref{fig:bazr} and \ref{fig:hslshs}. These figures show
plots of [Ba/Zr] vs. [Ba/Fe] and [hs/ls] vs. [hs/Fe],
respectively, for the whole sample, comprising all the metallicity
range for which information was available.


A large scatter can be seen in both plots (Fig. \ref{fig:bazr} and
Fig. \ref{fig:hslshs}) in the sense that a given [Ba/Zr] ratio (or
[hs/ls]) seems to correspond to a variety of [Ba/Fe] (or [hs/Fe])
values. In Fig. \ref{fig:bazr} there are two stars, \object{HD 26886}, a
mild barium star, and \object{HD 50082}, a barium star, that have a higher
[Ba/Zr] than the bulk of the other stars. Both were analyzed by
Liang et al.\@ (\cite{Li03}). Their larger [Ba/Zr] are due to a
smaller Zr abundance when compared to stars with similar Ba
abundance. Without these two stars the average [Ba/Zr] for the
mild barium stars is +0.00 $\pm$ 0.20 and +0.04 $\pm$ 0.12 for the
barium stars, or +0.06 $\pm$ 0.27 and +0.11 $\pm$ 0.20 including
them, respectively. In any case, as is clear from the plot, there
is no significant difference in the [Ba/Zr] ratio between barium
and mild barium stars. We have also plotted equivalent diagrams,
[Ba/Y] vs. [Ba/Fe], [La/Zr] vs. [La/Fe], and [La/Y] vs. [La/Fe],
and, similarly, no clear trend of an increasing ratio of a heavy
to a light s-process element with increasing excess of the
corresponding heavy s-process element could be discerned with any
statistical significance.

The number of points in Fig. \ref{fig:hslshs} is smaller, since
for some stars abundances for all the elements defining [hs] and
[ls] were not available. Nevertheless, in this plot there seems to
be a trend of increasing [hs/ls] with increasing [hs/Fe], when
considering only the peculiar giants. The barium star with smaller
[hs/ls] in the plot is HR 8204. Its low [hs/ls] is strongly
influenced by the high Sr overabundance, based on only one line,
which increases its [ls] mean. The average [hs/ls] for the mild
barium stars is $-$0.22 $\pm$ 0.16 and for the barium stars is
+0.01 $\pm$ 0.21, or +0.07 $\pm$ 0.14 without HR 8204. A possible
conclusion might be that, when one considers all the available
[X/Fe] ratios in composing the [hs] and [ls] means, and in spite
of large scatter, mild barium stars seem to have a slightly lower
level of neutron exposure than classical barium stars, even though
their metallicity range is exactly the same.

However, we still have to be very careful in drawing a conclusion. The
position of HR 8204 seems to indicate that it is possible to
produce a barium star with a [hs/ls] as low as that observed in
some mild barium giants. Moreover, the mild barium star with the
highest [Ba/Zr] in Fig. \ref{fig:bazr} is not included in Fig.
\ref{fig:hslshs}, due to the lack of the abundances of the
necessary elements, and therefore we cannot be sure there are not
mild barium stars with [hs/ls] ratios as high as the peak observed
for the barium giants. The addition of more points like those
could blur the weak correlation seen in Fig. \ref{fig:hslshs}.
Thus, although we have indications that the material in barium
stars was subject to a higher neutron exposure, a firm conclusion
should await an increase in the sample.

In the case mild barium and barium stars do share the same range
in neutron exposure their differences should be related to yet
another factor other than either neutron exposure, metallicity or
larger orbital period. The difference could, for example, be
connected to the mass range of the progenitors, and the associated
nucleosynthetic processes, and their ability to mix s-process
enriched material to the surface during the third dredge-up. It
could also be related to the convective mixing of the barium stars
themselves during the RGB, the first dredge-up, acting to dilute
the overabundances.

We do not have $^{12}$C/$^{13}$C or C and N abundances for the
sample stars, hence we cannot discuss their evolutionary status in
detail. However, in Fig. \ref{fig:hr} we see that both barium and
mild barium stars of our sample seem to share the same region in
the HR diagram and to have similar masses. A great difference in
the convective mixing efficiency in this restricted range of
stellar masses is not expected (Schaller et al.\@ \cite{Sc92}). We
cannot discard the possibility that the origin of the
differences lie in a complex combination of all these phenomena,
mixing, mass loss, and neutron exposure with some role played by
metallicity dependence or orbital separation.

A straightforward conclusion on the origin of the different
overabundances in barium and mild barium is not yet possible. We
discarded the hypothesis of different metallicities, at least in the range
of parameters defined by our sample, but we  also
showed that there seems to be a difference in the neutron exposure
range. Further observational work on extending the sample of
barium and mild barium stars with detailed abundance analysis is 
necessary as well as additional theoretical work on
formation scenarios for these systems.

\subsection{Copper in s-process enhanced stars}

The nucleosynthetic sites of Cu production are still poorly known.
Sneden et al.\@ (\cite{S91}) derived  Cu abundances for a large
sample of field and globular cluster stars and suggested that Cu
was mainly produced by the weak component of the s-process in
massive stars, with only a small contribution of type I supernovae
and of the main component of the s-process in intermediate mass
stars. Bisterzo et al.\@ (\cite{Bi04}) discuss a collection of Cu
abundances from the literature of a variety of systems, field
stars (from halo, thick and thin disk), bulge-like stars, globular
cluster stars and stars from dwarf spheroidal galaxies. They
conclude that the Cu behavior can be explained as the result of an
efficient weak s-process in massive stars. McWiliam \&
Smecker-Hane (\cite{MSH05}) show that this conclusion is
consistent with the Cu abundances in the Sagittarius dwarf
spheroidal and the chemical evolution scenario earlier proposed to
explain the Mn abundances of this system (McWilliam et al.\@
\cite{Mc03}).

On the other hand, Matteucci et al. (\cite{M93}) argue that a long
lived process, such as type Ia supernovae, is necessary to bring
the observations and theory of the Galactic chemical evolution of
Cu into agreement. Mishenina et al.\@ (\cite{M02}), by means of an
abundance analysis of a large sample of metal-poor halo and thick
disc stars, also conclude that the non-linear trend of [Cu/Fe]
with [Fe/H] is best explained if the bulk of Cu comes from
explosive nucleosynthesis in type Ia supernovae. Abundances of Cu
in $\omega$ Cen (Cunha et al.\@ \cite{Cu02}) and other globular
clusters (Simmerer et al.\@ \cite{Si03}) also seem to indicate
that Cu is mainly produced by type Ia supernovae.

Further investigation, both on theoretical yields and on the
observational behavior of Cu in a variety of systems, is still
needed in order to understand the nucleosynthetic origin of Cu. In
addition to this discussion, Castro et al.\@ (\cite{C99}) derived
Cu and Ba abundances for a sample of barium enhanced dwarfs from
the UMaG (Soderblom \& Mayor \cite{SM93}) and found the existence
of an anticorrelation between [Cu/Fe] and [Ba/Fe]. This
anticorrelation is reinforced by the Cu deficiency observed in two
yellow symbiotic stars that are s-process enhanced (Pereira \&
Porto de Mello \cite{P97}; Pereira et al.\@ \cite{P98}). This is a
possible indication that besides being built by the s-process, Cu
could also be a seed to the production of heavier elements. Its
depletion in s-process enriched stars could be a sign of
preferential use as seed. One of the goals of this work was to
verify whether this anticorrelation is also present in barium
stars.

\begin{figure}
\centering
\includegraphics[width=7cm,height=6cm]{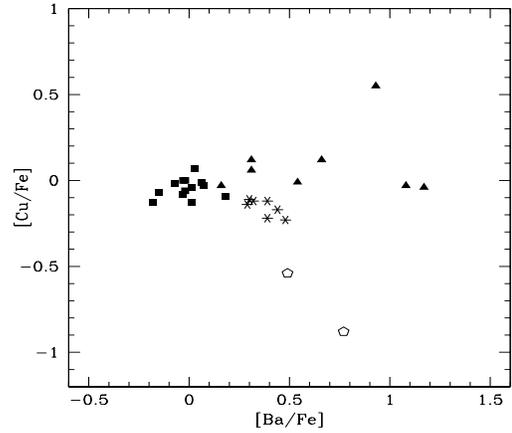}
\caption{Plot of [Cu/Fe] vs. [Ba/Fe]. In this figure the triangles
are the barium and mild barium stars analyzed in this work. The
squares are normal disc stars from Castro et al.\@ (\cite{C99}).
The asterisks are s-process enriched dwarfs of the UMaG also from
Castro et al.\@ (\cite{C99}). The open symbols are yellow
symbiotic stars from Pereira \& Porto de Mello (\cite{P97}) and
Pereira et al.\@ (\cite{P98}).}
\label{fig:cuba}
\end{figure}

Fig. \ref{fig:cuba} shows the plot of [Cu/Fe] vs. [Ba/Fe] for the
UMaG stars from Castro et al.\@ (\cite{C99}), the barium and mild
barium stars of this work, the two yellow symbiotic stars from
Pereira \& Porto de Mello (\cite{P97}) and Pereira et al.\@
(\cite{P98}) and a sample of normal disk stars also from Castro et
al.\@ (\cite{C99}). Barium and mild barium stars, however, do not
follow the anticorrelation. On the contrary, they seem to follow
the plateau defined by the normal disk stars. This result argues
that the observed depletion is not a common fact that extends to
all s-process enhanced stars. We note however that the origin of
this anticorrelation is still not clear and deserves further
investigation.

\begin{figure}
\centering
\includegraphics[width=7cm,height=6cm]{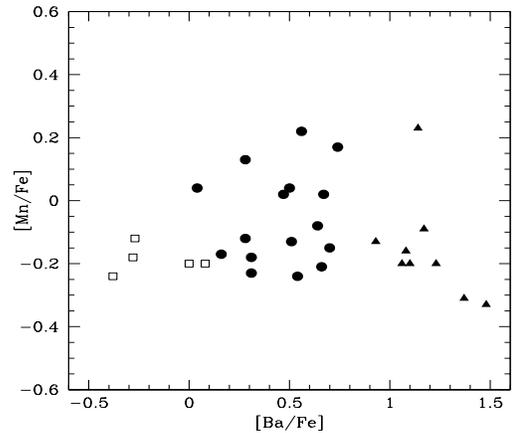}
\caption{Plot of [Mn/Fe] vs. [Ba/Fe]. In this figure the triangles are barium
stars, the circles are mild barium stars and the open squares are
normal giants.}
\label{fig:mnba}
\end{figure}

\begin{figure}
\centering
\includegraphics[width=7cm,height=6cm]{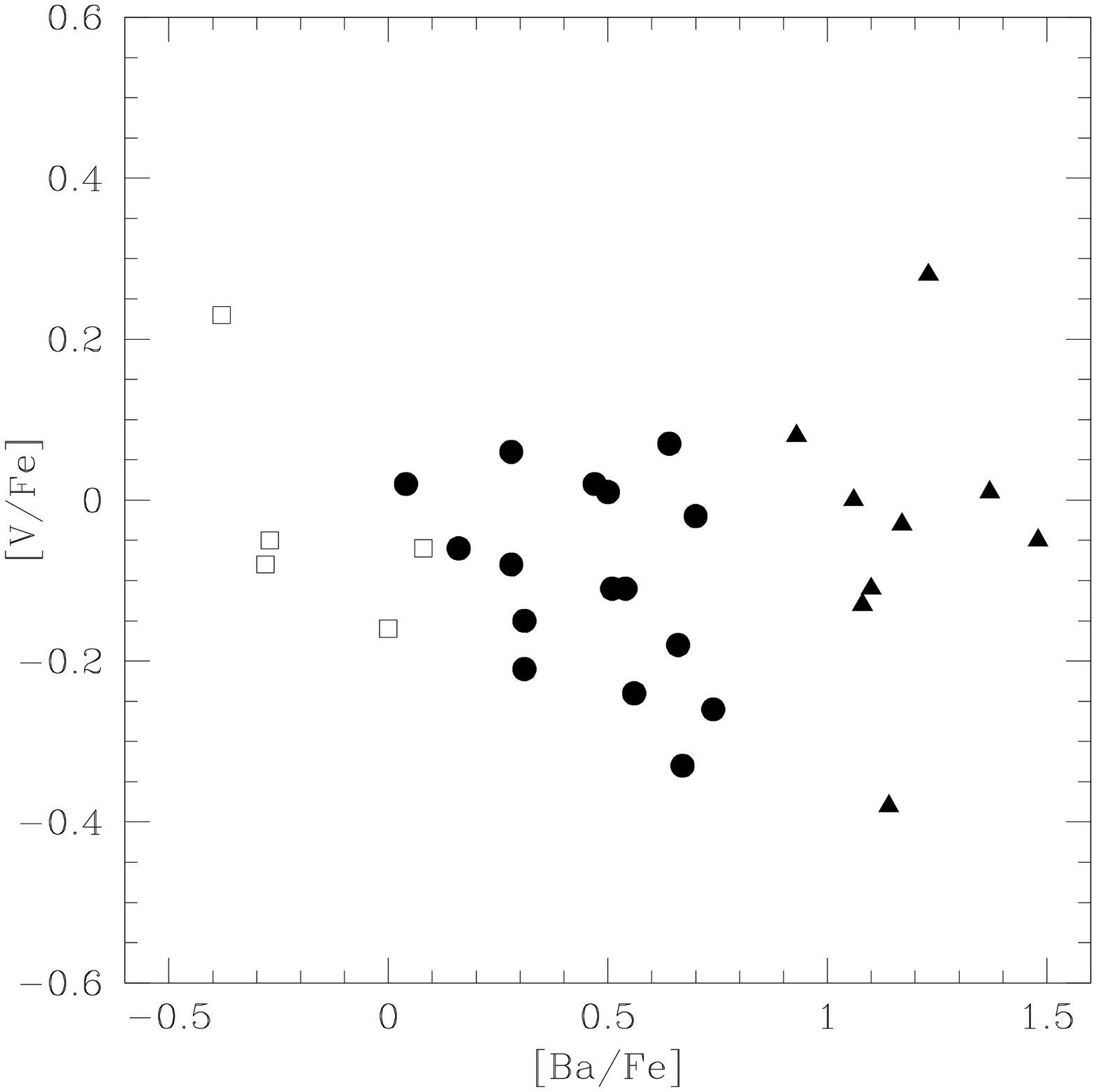}
\caption{Plot of [V/Fe] vs. [Ba/Fe]. Symbols are as in fig. 24.}
\label{fig:vba}
\end{figure}

\begin{figure}
\centering
\includegraphics[width=7cm,height=6cm]{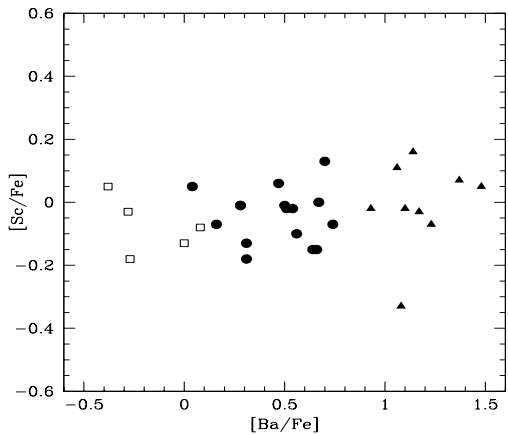}
\caption{Plot of [Sc/Fe] vs. [Ba/Fe]. Symbols are as in fig. 24.}
\label{fig:scba}
\end{figure}

The analysis of HR 6094, a s-process enhanced UMaG member (Porto de
Mello \& da Silva \cite{PS97}) also suggests a depletion of Mn and
excess of V and Sc. Neither of these effects is seen in the stars
we analyzed or in the data of the barium and mild barium stars we
collected from the literature. We show in figures \ref{fig:mnba},
\ref{fig:vba} and \ref{fig:scba} the plots of [Mn/Fe], [V/Fe] and
[Sc/Fe] vs. [Ba/Fe] respectively. In these the triangles are
barium stars, the circles are mild barium stars and the squares
normal giants. No clear-cut difference in Mn, V or Sc content is
apparent between the three group of stars, except for a larger
scatter of Mn and V in the chemically peculiar stars.


\section{Conclusions}

We carried out a detailed analysis of a sample of eleven barium
and mild barium stars and three normal giants using high
resolution, high signal to noise spectra. We determined
atmospheric parameters, ages, masses and abundances for
twenty-five elements; Na, Mg, Al, Si, Ca, Sc, Ti, V, Cr, Mn, Fe,
Co, Ni, Cu, Zn, Sr, Y, Zr, Ba, La, Ce, Nd, Sm, Eu and Gd.

We found the stars HR 649 and HR 1016 to be normal red giants, not
mild barium stars, in contrast to what was generally accepted in
the literature. Since there are other cases in the literature of
mild barium stars found to be misclassified (Smith \& Lambert
\cite{SL87}) we suggest that a verification of the peculiar status
of a large sample of tentative mild barium stars is very much
needed. With this aim we are currently analyzing a large sample of
mild barium stars, and the results will be presented in a
forthcoming paper.

The abundances of barium and mild barium stars were compared and
we found that there seems to be no difference in iron abundance
between them. The two groups seem have the same metallicity range
of normal disk giants and thus barium and mild barium stars seem
to be members of the same stellar population.

We found some indications that the material transferred onto
barium stars could have been subjected to a higher neutron
exposure than that accreted by the mild barium stars. More work,
however, is needed to confirm this result. The reasons for this
difference are not yet clear. Metallicity does not seem to be an
issue, but possibly higher neutron exposures are associated with
higher excesses of the heavier s-process elements. Parameters
which might be involved are the mass range of the former primaries
of these systems, or differences in the mixing processes of the
barium stars themselves.

A possible anticorrelation between [Cu/Fe] and [Ba/Fe] seen in
some s-process enriched stars (Castro et al.\@ \cite{C99}) was not
identified in the barium or mild barium stars of our sample. These
seem to follow the plateau defined by the normal disk stars. This
result argues that the observed depletion of Cu in some barium
enhanced stars is not a common feature that extends to all
s-process enhanced stars. The origin of the anticorrelation
deserves further investigation. Possible similar effects for Sc, V
and Mn were not identified either in the barium and mild barium
stars of our sample.

\begin{acknowledgements}

This paper is based on the senior thesis of RS. GFPM would like to
thank Verne V. Smith, Roberto Gallino and Dinah M. Allen for
helpful and stimulating discussions on the subject of barium
stars. RS acknowledges a FAPERJ fellowship (E-26/150618/2000)
during the development of this work, and later financial
support from CAPES and FAPESP. GFPM acknowledges financial support
from FAPERJ (grant APQ1/26/170687/2004), CNPq/Conte\'udos Digitais
(grant 552331/01-5) and from CNPq/Instituto do Mil\^enio,
620053/2001-1. L. da S. thanks the CNPq, Brazilian Agency, for the
grant 30137/86-7.
%FAPESP 00/06769-4 and PRONEX/FINEP 41.96.0908.00.

\end{acknowledgements}

%--------------------------------- References -------------
 
%
\Online
\begin{longtable}{cccccccccc}
\caption{\label{tab:le} Equivalent widths for the stars HR440, HR649, HR1016, HR1326, HR2392 and HR4608. The derived log $gfs$ are also listed.} \\
\hline\hline
$\lambda$ (\AA) & Elem. & $\chi$ (eV) & log $gf $ & HR440 & HR649 & HR1016 & HR1326 & HR2392 & HR4608 \\
\hline
\endfirsthead
\caption{continued.} \\
\hline\hline
$\lambda$ (\AA)  &    Elem. & $\chi$ (eV) &    log $gf $ &  HR440 &  HR649  &  HR1016 &  HR1326 &  HR2392  &  HR4608 \\
\hline
\endhead
\hline
\endfoot
6154.230 & NaI & 2.10 & -1.179 &  60.4 &  75.5 &  73.5 &  89.1 &  81.9 &  55.5 \\
6160.753 & NaI & 2.10 & -0.905 &  83.8 &  98.2 &  92.7 & 107.1 & 105.4 &  76.6 \\
4571.102 & MgI & 0.00 &   --   & 194.5 &   --  & 193.6 &   --  & 259.6 & 186.8 \\
4730.038 & MgI & 4.34 &   --   & 107.5 & 104.8 & 108.5 & 126.2 & 141.3 & 105.2 \\
5711.095 & MgI & 4.34 &   --   & 135.1 & 133.7 & 127.6 & 148.2 & 142.8 & 125.0 \\
5785.285 & MgI & 5.11 & -1.615 &  76.9 &  78.4 &  76.5 &  91.6 &  87.0 &  62.9\\
6696.032 & AlI & 3.14 & -1.370 &  66.7 &  58.4 &  58.4 &  91.5 &  67.1 &  56.3\\
6698.669 & AlI & 3.14 & -1.659 &  43.9 &  38.3 &  40.0 &  69.1 &  65.2 &  38.4\\
\hline
\end{longtable}
\begin{longtable}{cccccccccc}
\caption{\label{tab:le2} Equivalent widths for the stars $\epsilon$ Vir, HR5058, HR5802, HR7321, HR8115, HR8204, HD205011 and HR8878.} \\
\hline\hline
$\lambda$ (\AA) & Elem. & $\epsilon$ Vir & HR5058 & HR5802 & HR7321 & HR8115 & HR8204 & HD205011 & HR8878 \\
\hline
\endfirsthead
\caption{continued.} \\
\hline\hline
$\lambda$ (\AA) & Elem. & $\epsilon$ Vir & HR5058 & HR5802 & HR7321 & HR8115 & HR8204 & HD205011 & HR8878 \\
\hline
\endhead
\hline
\endfoot
6154.230 & NaI &  94.4 &  87.1 &  81.0 &  70.7 &  95.1 &  67.7 &  80.7 &  63.5 \\
6160.753 & NaI & 113.1 & 122.6 & 102.3 &  92.7 & 114.7 &  93.4 & 101.8 &  87.1 \\
4571.102 & MgI & 204.6 & 244.0 & 201.8 & 213.8 & 208.9 & 214.8 & 225.0 &   --  \\
4730.038 & MgI & 122.9 & 143.3 & 125.9 & 119.2 & 120.5 & 133.0 &   --  & 112.4 \\
5711.095 & MgI & 145.0 & 140.4 & 139.9 & 134.2 & 140.9 & 132.2 & 136.4 & 145.0 \\
5785.285 & MgI &  88.2 &  92.8 &  81.9 &  81.3 &  86.9 &  75.3 &  84.0 &  71.4 \\
6696.032 & AlI &  74.7 &  77.0 &  71.7 &  47.3 &  72.1 &   --  &  74.3 &  80.9 \\
6698.669 & AlI &  54.5 &  70.6 &  54.0 &  66.5 &  55.6 &  69.0 &  59.4 &  55.2 \\
\hline
\end{longtable}
\begin{longtable}{cccccc}
\caption{\label{tab:gfs} Log $gfs$ for each component of the lines with hyperfine structure.} \\
\hline\hline
\endfirsthead
\caption{continued.} \\
\hline\hline
\endhead
\hline
\endfoot
4739.113 & MnI   &    5413.684 & MnI   &  5420.350 & MnI  \\
\hline
$\lambda$ (\AA) &  log gf  &$\lambda$ (\AA) &  log gf  & $\lambda$ (\AA) &  log gf  \\
4739.099 & -1.269 &  5413.613 & -1.716 &  5420.277 & -2.121 \\
4739.113 & -1.410 &  5413.653 & -0.921 &  5420.301 & -2.043 \\
4739.126 & -1.565 &  5413.679 & -1.169 &  5420.334 & -2.903 \\
4739.145 & -1.122 &  5413.714 & -0.805 &  5420.376 & -1.796 \\
4739.167 & -2.472 &  5413.722 & -1.265 &  5420.429 & -1.708 \\
\hline\hline
5432.548 & MnI   &    5537.765 & MnI  &  6013.497 & MnI  \\
\hline
\end{longtable}

\end{document}